\documentclass[aps,showpacs,eqsecnum,10pt]{revtex4}
\usepackage{graphicx}
\usepackage{amsmath}
\usepackage{amssymb}

\newcommand{\be}{\begin{equation}}
\newcommand{\ee}{\end{equation}}
\newcommand{\ba}{\begin{eqnarray}}
\newcommand{\ea}{\end{eqnarray}}

\def\pref#1{(\ref{#1})}

\begin{document}

\title{Magnetic oscillations in planar
systems with the Dirac-like spectrum of quasiparticle excitations}

\author{S.G.~Sharapov$^{1}$}
\email{sharapov@isiosf.isi.it}
\thanks{Present address: Institute for Scientific Interchange, via Settimio Severo 65,
I-10133 Torino, Italy}
\author{V.P.~Gusynin$^{2}$}
\thanks{On leave of absence from Bogolyubov Institute for Theoretical Physics,
        Metrologicheskaya Str. 14-b, Kiev, 03143, Ukraine}
\email{vgusynin@bitp.kiev.ua}

\author{H.~Beck$^1$}%
\email{Hans.Beck@unine.ch}
\homepage{http://www.unine.ch/phys/theocond/}

\affiliation{
        $^1$Institut de Physique,
        Universit\'e de Neuch\^atel, 2000 Neuch\^atel, Switzerland\\
        $^2$Department of Applied Mathematics, University of Western
        Ontario, London, Ontario N6A 5B7, Canada}

\date{\today }

\begin{abstract}
The quantum magnetic oscillations are studied for planar condensed
matter systems with a linear, Dirac-like spectrum of quasiparticle
excitations. We derive analytical expressions for magnetic
oscillations (de Haas -- van Alphen effect) in the density of
states,  thermodynamic potential, magnetization, and chemical
potential both for zero and finite temperatures, and in the
presence of scattering from impurities. We discuss also a
possibility of using magnetic oscillations for detection of a gap
that may open in the spectrum of quasiparticle excitations due to
a nontrivial interaction between them.
\end{abstract}




\maketitle

\section{Introduction}
Magnetic oscillations (MO) in metals \cite{Shoenberg.book} proved
to be a powerful tool for investigation of the shape of their
Fermi surface. The oscillations of magnetization were predicted by
Landau in 1930 \cite{Landau:1930}. The experimental discovery of
MO came soon: Shubnikov and de Haas (SdH) found oscillations of
electrical conductivity in Bi crystals, and later de Haas and van
Alphen (dHvA) discovered the oscillations of magnetization.
Despite more than 70 years old history of MO they continue to
attract attention of both experimentalists and theorists. The
recent experimental studies are mostly focused on
quasi-two-dimensional (quasi-2D) organic conductors and
superconductors (for a review, see e.g.
\cite{Singleton:2000:RPP}). The ultimate hope is that better
understanding of the 2D and quasi-2D organic systems can also
contribute to the studies of high-temperature superconductors that
have a similar layered structure. As suggested recently in
Ref.~\cite{Yasui:2002:PRB}, dHvA experiment can be used as a probe
to detect band- and/or angle-dependent gap amplitudes of type-II
superconductors.

It turns out that these experimental advances demand also further
development of the MO theory. While MO observed in 3D metals are
well described by Lifshits-Kosevich (LK) \cite{Lifshits:1956:JETP}
(for dHvA effect) and Adams-Holstein \cite{Adams:1959:JPCS} (for
SdH effect) theories, there is no commonly accepted and used
theory for MO in 2D and quasi-2D materials \cite{Champel:2001:PM}.
As discussed in Ref.~\cite{Champel:2001:PM}, there are two
essential
features of dHvA effect in 2D case that differ it from 3D case: \\
(1) The sharp saw-tooth like shape of the oscillations seen in the
low temperature $T \ll \omega_L$ and high purity $\Gamma \ll
\omega_L$ regions. Here, $\omega_L$ is the distance between Landau
levels and $\Gamma$ is the width of Landau's levels due to
impurity scattering.\\
(2) Due to the Landau quantization of the 2D kinetic energy in the
magnetic field, both the density of electrons, $n$, and the
chemical potential, $\mu$, cannot be fixed simultaneously as the
magnetic field is changed. Depending on the
physical situation, two extreme limits are usually considered:\\
\indent (i) the limit of fixed and field independent chemical
potential $\mu$, which can be represented by a grand canonical
ensemble. This corresponds to the case originally studied by
Lifshits and Kosevich \cite{Lifshits:1956:JETP} in 3D.\\
\indent (ii) the limit of fixed density of carriers $n$, in which
a strong dependence of the chemical potential $\mu$ on the
magnetic field that has to be taken into account by considering an
equation for $\mu$ that leads to the chemical potential
oscillations. This equation is crucial in 2D case even for the
large Fermi energy, $\epsilon_F \gg \omega_L$ since despite
smallness of the chemical potential shift, $\delta\mu\sim\omega_L$
due to the magnetic field compared to $\epsilon_F$, it can change
the phase of the magnetization MO (in 3D the oscillating part
$\delta\mu\sim\omega_L(\omega_L/\epsilon_F)^{1/2}$, so  that for
$\epsilon_F \gg \omega_L$ it does not alter the MO with a high
accuracy). Nevertheless, under certain conditions discussed in
Refs.~\cite{Champel:2001:PM,Champel:2001:PRB} the oscillations of
the chemical potential are small and one may still assume that the
chemical potential $\mu$ coincides with the Fermi energy
$\epsilon_F$ as in the limit (i).

On the other hand, in condensed matter systems with a linear,
``relativistic" spectrum of quasiparticle excitations, the
conditions favorable for MO seem to persist to rather small
magnetic fields and high temperatures \cite{Abrikosov:1998:PRB}.
This is particularly true for 2D systems with the Dirac
quasiparticle spectrum.

There is a wide variety of planar condensed matter systems that in
the low-energy limit have a linear dispersion law of the
quasiparticle excitations. In particular, the studies of
high-temperature cuprate superconductors inspired a big interest
in the so-called nodal excitations that are the gapless fermion
excitations associated with zeros of the gap function. Depending
on the physical origin of the gap, one can consider rather
different physical situations that are in general described by
different theoretical models. For example, if the gap opens  due
to the anisotropic electron-hole pairing, which is one of the
possible states of the electron system with a half-filled band and
nested Fermi surface, this corresponds to a 2D orbital
antiferromagnet (OAF) or staggered flux state
\cite{Halperin:1968:SSP,Marston:1989:PRB,Nersesyan:1989:JLTP}.

There is a consensus that the superconducting state in cuprates
has a $d$-wave superconducting energy gap, with nodes along the
diagonals of the Brillouin zone \cite{Tsuei:2000:RMP}. A linearization
of the Bogolyubov spectrum of quasiparticles around four nodes
on the Fermi surface leads to another realization of gapless fermion
excitations.
Although from physical point of view the
OAF state seems to be more complicated than
the $d$-wave superconducting state, because the former is
characterized by nonzero local currents violating time-reversal
and translational symmetries, the Dirac description of OAF state
turns out to be simpler since an external electromagnetic field enters
into the theory in the same way as in QED$_{2+1}$.

Carbon-based materials, e.g. pyrolytic graphite and carbon
nanotubes present one more example of the planar systems with the
relativistic dispersion law
\cite{Semenoff:1984:PRL,Vincenzo:1984:PRB,Gonzales:1993:NP}. From
an experimental point of view these materials are probably the
most promising because the quality of available samples already
allows one to observe SdH effect \cite{Uji:1998:PB,Wang:2003:PLA}
and to study quantum  effects that are due to a high magnetic
field \cite{Kopelevich:2003:PRL} (see Ref.~\cite{Kopelevich:2002}
for a review).

The relativistic theories in an external magnetic field have been
the subject of  research in quantum field theory for many years
(for reviews see Ref.~\cite{Dittrich.book}). The extraction of MO
from the general expressions presents a rather subtle and
interesting problem that was investigated in
Refs.~\cite{Elmfors:1993:PRL,Vshivtsev:1995:JETP,Vshivtsev:1996:JETP,
Cangemi:1996:AP,Inagaki:2003}. However these studies of the MO in
QED are mostly concentrated on the field theoretical aspects.

The purpose of the present paper is to make a systematic study of
the MO in QED$_{2+1}$ devoting special attention to the link with
planar condensed matter systems and to quantities that are
particularly important for condensed matter theory, e.g. density
of states (DOS), thermodynamic potential and magnetization. To
make the theory more realistic we also include into the model the
effect of scattering from impurities, by considering Landau levels
with field and temperature independent width.

We begin by presenting in Sec.~\ref{sec:model} three
2D models  mentioned above that can be studied using the same effective
relativistic Lagrangian written in Sec.~\ref{sec:Dirac} in the presence of
an external magnetic field.
The Green's function necessary for subsequent calculations
is introduced in Sec.~\ref{sec:GF} and the DOS oscillations are studied in
Sec.~\ref{sec:DOS} both with and without scattering from impurities.
The general representation for the thermodynamic potential in
terms of the DOS and its link to the corresponding nonrelativistic
potential are considered in Sec.~\ref{sec:Omega.general}. In
Sec.~\ref{sec:Omega0.Gamma=0} we derive the expression for the
thermodynamic potential at $T=0$ in the absence of impurities with
explicitly extracted MO (the calculational details and an
alternative representation for the thermodynamic potential in
terms of the generalized zeta function are given in
Appendices~\ref{sec:A} and \ref{sec:B}). In Sec.~\ref{sec:gap}
and at the end of Sec.~\ref{sec:Omega.T} we
discuss a possibility of detecting a gap that may open in the
spectrum of one of the systems discussed above. The analytical
expression for the magnetization at $T =0$ in the absence of
impurities written in terms of Bernoulli polynomials is given in
Sec.~\ref{sec:magnetization}. In Sec.~\ref{sec:Omega.Gamma-T} we
generalize the results obtained in the previous sections
to the presence of impurities (Secs.~\ref{sec:number.Gamma} and \ref{sec:Omega.Gamma}
with the calculational details presented in Appendix~\ref{sec:C})
and in Sec.~\ref{sec:Omega.T} for nonzero temperature. In
Conclusions, Sec.~\ref{sec:conclusions}, we give a concise summary
of the obtained results.

\section{Model}
\label{sec:model}

There are many planar condensed matter models that in the low
energy sector can be reduced to QED$_{2+1}$ form and here we briefly
describe some of them underlining the main assumptions leading
to the Dirac-like form of the effective Hamiltonians.

\subsection{Hamiltonian of $d$-density wave state}

Due to $d_{x^2 - y^2}$ momentum dependence of the gap
in the OAF states it is also called $d$-density wave (DDW)
state \cite{Chakravarty:2001:PRB}.
The mean-field phenomenological Hamiltonian describing this state
can be written as
\cite{Nersesyan:1989:JLTP,Yang:2002:PRB}
\begin{equation}
\label{Hamiltonian.DDW.matrix}
H^{\mathrm{DDW}}  = \sum_{\sigma = \uparrow, \downarrow}
\int_{\mathrm{RBZ}} \frac{d^2 k}{(2 \pi)^2}
\chi_{\sigma}^{\dagger}(\mathbf{k}) \left[ \varepsilon(\mathbf{k}) \sigma_3
- \mu \hat{I} - D(\mathbf{k}) \sigma_2 \right] \chi_{\sigma}( \mathbf{k}),
\end{equation}
where the spinors
\begin{equation}
\label{spinors}
\chi_\sigma(\mathbf{k}) = \left( \begin{array}{c}
c_{\sigma}(\mathbf{k}) \\
c_{\sigma}(\mathbf{k} + \mathbf{Q})
\end{array} \right), \qquad
\chi_\sigma^{\dagger}(\mathbf{k}) = \left( \begin{array}{cc}
c_{\sigma}^{\dagger}( \mathbf{k})
\quad c_{\sigma}^{\dagger}(\mathbf{k + Q})
\end{array} \right),
\end{equation}
are composed from creation and annihilation operators
$c_{\sigma}^{\dagger}(\mathbf{k})$, $c_\sigma(\mathbf{k})$ for
momentum $\mathbf{k}$ and spin $\sigma$, the single particle
energy is $\varepsilon({\mathbf{k}}) = - 2 t (\cos k_x a + \cos
k_y a)$ with $t$  being the hopping parameter, $\mu$ is the
chemical potential, $D(\mathbf{k}) = \frac{D_0}{2}(\cos k_x a -
\cos k_y a )$ is the $d$-density wave gap and $\mathbf{Q} =
(\pi/a, \pi/ a)$ is the wave vector at which the density-wave
ordering takes place. The integral is over the reduced Brillouin
zone $\mathrm{RBZ}$ and $\sigma_i$ are Pauli matrices. Throughout
the paper $\hbar = c = k_{B} = 1$ units are chosen, unless
stated explicitly otherwise.

This Hamiltonian describes the excitations with the spectrum
$E(\mathbf{k}) =  - \mu \pm \sqrt{\varepsilon^2(\mathbf{k}) +
D^2(\mathbf{k})}$. Linearizing the spectrum about the four nodes
$\mathbf{N}_i = (\pm \pi/2a, \pm \pi/2a )$ with $i=1,\ldots ,4$ at
half-filling
($\mu =0$) one obtains $E(\mathbf{k}) = - \mu \pm  \sqrt{v_F^2 k_x^2 +
v_D^2 k_y^2}$, where the Fermi velocity calculated for
half-filling $v_F = |\partial \varepsilon(\mathbf{k})/\partial
\mathbf{k} |_{\mathbf{k} = \mathbf{N}} | = 2 \sqrt{2} t a$, the
DDW gap velocity $v_{D} = |\partial D(\mathbf{k})/\partial
\mathbf{k} |_{\mathbf{k} = \mathbf{N}} | = \frac{1}{\sqrt{2}} D_0
a$, and the momenta $k_x$ and $k_y$ are given in the local nodal
coordinate system (see Fig.~2 of Ref.~\cite{Yang:2002:PRB}).

Using this linearized spectrum and inserting the vector potential
in a way that preserves the charge conservation for the original
Hamiltonian (\ref{Hamiltonian.DDW.matrix}), one can arrive
\cite{Sharapov:2003:PRB} at the following Dirac Lagrangian
describing the quasiparticles in the vicinity of one of the nodes,
e.g. $\mathbf{N}_{i=1}=(\pi/2a, \pi/2a)$
\begin{equation}
\label{Dirac.conjugated.Lagrangian}
\mathcal{L} =  \bar{\chi}_\sigma^i (x) i \tilde{\gamma}^{\nu} D_{\nu}\chi_\sigma^i (x),
\qquad x = (t, \mathbf{r}),
\end{equation}
where $\bar{\chi}_\sigma^i (x) = \chi_\sigma^{i \dagger}(x) \sigma_1$
is the  Dirac conjugated spinor, $i$ labels the node $\mathbf{N}_i$,
the covariant derivatives  $D_\nu$ are
\begin{equation}
\label{long.derivative}
D_{\nu} =
\begin{cases}
\hbar \partial_t  - i e A_{0}(x), &      \nu =0,\\
v_{F} \left(\hbar \partial_x - i \frac{e}{c} A_{1}(x) \right), &  \nu =1, \\
v_{\Delta} \left( \hbar \partial_y - i \frac{e}{c} A_{2}(x)\right), &  \nu =2,
\end{cases}
\end{equation}
and  the $\gamma$ matrices are
\begin{equation}
\label{DDW.gamma}
\tilde{\gamma}^{\nu} = (\sigma_1,  - i \sigma_2, i \sigma_3),
\qquad \{\tilde{\gamma}^\mu, \tilde{\gamma}^{\nu}\} = 2 \hat{I}_2 g^{\mu \nu}, \qquad
g^{\mu \nu} = \mbox{diag}(1,-1,-1), \qquad \mu, \nu = 0,1,2.
\end{equation}
The chemical potential $\mu$ can be introduced in the
Lagrangian (\ref{Dirac.conjugated.Lagrangian})
by choosing $A_0 = \mu/e$ and the sum over the spin components in
Eq.~(\ref{Dirac.conjugated.Lagrangian}) can be regarded as an additional
flavor index.

In 2+1 dimensions, there are two inequivalent representations
of Dirac algebra (see e.g.~\cite{Jackiw:1981:PRD}):
\begin{subequations}
\label{gamma.standard}
\begin{align}
& \hat{\gamma}^0 = \sigma_3, \qquad \hat{\gamma}^1 = i \sigma_1,
\qquad \hat{\gamma}^2 = i \sigma_2, \label{a.gamma} \\
& \hat{\gamma}^0 = - \sigma_3, \qquad \hat{\gamma}^1 = - i \sigma_1,
\qquad \hat{\gamma}^2 = - i \sigma_2, \label{b.gamma}
\end{align}
\end{subequations}
which correspond to right and left handed coordinate systems. As
one can check, our $\gamma$-matrices (\ref{DDW.gamma}) correspond
to the representation (\ref{b.gamma}) if one makes a unitary
transformation $\hat{\chi}^1 = U \chi^1$, $\hat{\gamma}^\mu = U
\tilde{\gamma}^{\mu} U^{-1}$ with $U = (1/2)(\hat{I}_2 - i
\sigma_1 - i \sigma_2 +i\sigma_3)$ Since the physical properties
of the system depend only on the algebra that these matrices obey,
one can directly work with the more commonly used representation
(\ref{gamma.standard}) instead of (\ref{DDW.gamma}).

Dealing with (\ref{Dirac.conjugated.Lagrangian})
one should not forget that all physical quantities involve
the summation over the 4 nodes present in the original
Hamiltonian (\ref{Hamiltonian.DDW.matrix}).
In fact, only two neighboring nodes, e.g. $\mathbf{N}_{i=1} =(\pi/2a, \pi/2a)$
and $\mathbf{N}_{i=2}=(\pi/2a, - \pi/2a)$ are nonequivalent in RBZ and
the corresponding local nodal coordinate systems
are related to each other by parity transformation
$(k_x, k_y) \to (k_x, -k_y)$ with the simultaneous interchange
$v_F \leftrightarrow v_D$.
Thus the sum over nonequivalent nodes
can be taken into account by doubling spinors
\be
\Upsilon_{\sigma} =
\begin{pmatrix}
\psi^1_\sigma\\
\psi^2_\sigma
\end{pmatrix}
\ee
and using the reducible representation of $\gamma$-matrices,
$\gamma^{\mu} = (\sigma_3, i \sigma_1, i \sigma_2) \otimes \sigma_3$:
\begin{equation}
\label{gamma.matrices}
\gamma^{0} =
\begin{pmatrix}
\sigma_3 & 0 \\ 0 &-\sigma_3
\end{pmatrix}, \qquad
\gamma^{1} =
\begin{pmatrix}
i \sigma_1 & 0 \\ 0 &-i\sigma_1
\end{pmatrix}, \qquad
\gamma^{2} =
\begin{pmatrix}
i \sigma_2 & 0 \\ 0 &-i\sigma_2
\end{pmatrix}
\end{equation}
with the following Lagrangian density
\begin{equation}
\label{DDW.Dirac.reducible}
\mathcal{L} =  \bar{\Upsilon}_\sigma (x) i \gamma^{\nu} D_{\nu} \Upsilon_\sigma (x).
\end{equation}

\subsection{Hamiltonian of $d$-wave superconducting state}

In contrast to the hypothetical DDW state, $d$-wave
superconductivity ($d$SC) is observed in cuprate superconductors
\cite{Tsuei:2000:RMP}. It can be described by the Bogolyubov -
de Gennes Hamiltonian
\be
\label{Hamiltonian.BdG}
\begin{split}
H_{\mathrm{BdG}}&  = \int d^2 r \Psi^{\dagger} (t, \mathbf{r}) \tau_3
\left[ \varepsilon \left(- i \nabla - \tau_3 \frac{e}{c}\mathbf{A}(\mathbf{r})\right)
- \mu \right]
\Psi (t, \mathbf{r}) \\
& -  \int d^2 r_1 \int d^2 r_2
\left[ \Delta^{\dagger}(t, \mathbf{r}_1; \mathbf{r}_2)
\Psi^{\dagger}(t, \mathbf{r}_1) \tau_{-} \Psi(t, \mathbf{r}_2) +
\Psi^{\dagger}(t, \mathbf{r}_1) \tau_{+} \Psi(t, \mathbf{r}_2)
\Delta(t, \mathbf{r}_1; \mathbf{r}_2)
\right],
\end{split}
\ee where $\Psi$ and $\Psi^{\dagger} = (\psi^{\dagger}_{\uparrow},
\psi_{\downarrow})$ are the standard Nambu spinors,
$\varepsilon(\mathbf{p})$ is the single particle energy,
$\tau_{\pm} = (\tau_1 \pm i \tau_2)/2$ and $\Delta(t,
\mathbf{r}_1; \mathbf{r}_2)$ is the bilocal gap operator the
Hermitian conjugate of which includes the transpose in the
functional sense, i.e. $\Delta^{\dagger}(t, \mathbf{r}_1;
\mathbf{r}_2) \equiv [\Delta(t, \mathbf{r}_2;
\mathbf{r}_1)]^{\ast}$. The nontrivial dependence of  $\Delta(t,
\mathbf{r}_1; \mathbf{r}_2)$ on the relative coordinate
$\mathbf{r}_1 - \mathbf{r}_2$ is supposed to describe $d$-wave
pairing state. Further simplification of the pairing term of
Eq.~(\ref{Hamiltonian.BdG}) is possible if one assumes that the
amplitude of the pairing term is a constant with respect to $t$
and the center of mass coordinate $\mathbf{R} =
(\mathbf{r}_1+\mathbf{r}_2)/2$. For cuprates this assumption may
well be justified even above the critical temperature, $T_c$.
Below $T_c$ in the vortex state this assumption corresponds to
neglecting vortex core contributions and considering pure ``phase
vortices''. Since we deal with the amplitude of the bilocal
complex field $\Delta(t, \mathbf{r}_1, \mathbf{r}_2)$, it is
natural to consider also its phase $\theta(t, \mathbf{R})$.  There
are, however, some subtleties in writing the continuum version of
the pairing term of (\ref{Hamiltonian.BdG}) in an explicitly gauge
invariant form that can be solved by choosing an appropriate form
of the differential operator for $\hat{\Delta}$ (see
Refs.~\cite{Vafek:2001:PRB,Ye:2001:PRL}). Since we look for the
low-energy quasiparticle excitations near four gap nodes, we write
the linearized Hamiltonian for one of these nodes as (see e.g.
\cite{Vafek:2001:PRB,Yang:2002:PRB})
\begin{equation}
\label{Hamiltonian.SC}
H_{dSC} = \Psi^{\dagger} (x)\left[
i v_F \tau_3 \left(\partial_x - i \frac{e}{c} \tau_3 A_x \right)
+ i v_{\Delta} \tau_+ e^{i \theta(x)/2} \partial_y e^{i \theta(x)/2} +
i v_{\Delta} \tau_- e^{-i \theta(x)/2} \partial_y e^{-i \theta(x)/2}
\right] \Psi (x),
\end{equation}
where $v_{F}$ is the Fermi velocity, $v_{\Delta} = |\partial
\Delta(\mathbf{k})/\partial \mathbf{k} |_{\mathbf{k} = \mathbf{N}}
|$ is the gap velocity defined by the slope of the superconducting
gap $\Delta(\mathbf{k})$. Both velocities are calculated in the
nodal points on the true Fermi surface of the system, but not at
half-filling as for the DDW case. The linearization puts
restrictions on the domain of validity of the Hamiltonian
(\ref{Hamiltonian.SC}) which, however, remains rather wide
\cite{Simon:1997:PRL}.

Finally, introducing the Dirac conjugated spinor $\bar{\eta}^{i} =
\Psi^{i \dagger} \tau_2$ we can obtain from the Hamiltonian
(\ref{Hamiltonian.SC}) a Lagrangian similar to
Eq.~(\ref{Dirac.conjugated.Lagrangian}), but with the vector
potential that couples only with the $v_F$ term. Again making, if
necessary, a unitary transformation, one can combine spinors
$\eta^{i}$ originating from two opposite nodes into one
four-component spinor.

Thus in the absence of the field the low lying quasiparticle excitations
in the $d$SC are described by the relativistic dispersion law
\begin{equation}
E(\mathbf{k}) = \pm \sqrt{v_F^2 k_x^2 + v_{\Delta}^2 k_y^2},
\end{equation}
where the momenta $k_x$ and $k_y$ are given in the local nodal coordinate
system.
Experimental values of  $v_F$ and $v_{\Delta}$ for cuprates can be found in
Ref.\cite{Chiao:2000:PRB}, for example,
the value of $v_{F} \sim 2.5 \times 10^5 m/s$
and the anisotropy of the Dirac cone $v_{F}/v_{\Delta}$ varies
between 10 and 20 depending on doping and compound.

There are two important differences between DDW and $d$SC
Hamiltonians. Making a gauge transformation that removes the phase
from the last term of  Eq.~(\ref{Hamiltonian.SC}), we observe that
quasiparticles in superconductor do not couple simply to the
vector potential $\mathbf{A}$ corresponding to the magnetic field
$\mathbf{B} = \mathbf{\nabla} \times \mathbf{A}$, but to the
supercurrent $\mathbf{\nabla} \theta - (2 e/\hbar c )\mathbf{A}$,
where $\theta$ is the phase of the superconducting order
parameter. This difference between DDW and $d$SC cases leads to a
conclusion that in an external magnetic field the nodal
quasiparticles of DDW state form Landau levels
\cite{Nguyen:2002:PRB}, while in  $d$SC state Landau levels are
strongly mixed \cite{Franz:2000:PRL}. Nevertheless we presented
$d$SC Hamiltonian (\ref{Hamiltonian.SC}) here due to its practical
importance and the fact that the simplified physical picture based
on the Landau levels \cite{Gor'kov:1998:PRL,Anderson:1998} may
become relevant for higher energies \cite{Kopnin:2000:PRB} and can
be regarded as the first approximation to a more complicated case
of the vortex state in $d$-wave superconductors.

The second difference comes from the fact that for DDW state
chemical potential $\mu$ is explicitly present in the Lagrangian
(\ref{Dirac.conjugated.Lagrangian}), while for $d$SC state it was
absorbed in the definition of the Fermi velocity $v_F$ on the
Fermi surface. The origin of this difference can be traced back to
the different structures of the Hamiltonians
(\ref{Hamiltonian.DDW.matrix}) and (\ref{Hamiltonian.BdG}). One
can also say that in $d$SC state the chemical potential of nodal
quasiparticles is zero, so that when the applied field is changed
the corresponding Landau levels (even if they were formed) cannot
cross the chemical potential and produce dHvA effect
\cite{Gor'kov:1998:PRL}.

\subsection{Layered graphite}

The semimetallic energy band structure of a single graphene sheet
has the conduction and valence $\pi$ bands with the energy
dispersion \cite{Saito:book} \be E(k_x, k_y) = \pm t \sqrt{1+ 4
\cos \frac{\sqrt{3} k_x a}{2} \cos \frac{k_y a}{2} + 4 \cos^2
\frac{k_y a}{2}}, \ee where $t \approx 3 eV$, $a = \sqrt{3} a_{CC}
=2.46 \mbox{\AA}$ is the lattice constant of two dimensional
graphite, $a_{CC}$ is the distance between two nearest carbon
atoms. These two bands touch each other and cross the Fermi level
in six K points located at the corners of the hexagonal 2D
Brillouin zone, but only two of them, for example, $\mathbf{K} =
(2 \pi/a)(1/\sqrt{3}, 1/3)$ and $\mathbf{K}^\prime = (2 \pi/a)(0,
2/3)$ are inequivalent due to the periodicity of the Brillouin
zone. The low-energy excitations can be studied by taking the
continuum limit $a \to 0$ at any two independent K points labelled
as $j =1,2$. They have a linear dispersion $E_k = \pm v_{F} k$
with $v_{F} = (\sqrt{3}/2) t a \approx  9.7 \times 10^5m/s$. These
excitations can be formally described by a pair of two-component
(Weyl) spinors $\psi_{j \sigma}$, which are composed of the Bloch
states residing on the two different sublattices of the
bi-particle hexagonal lattice of the graphene sheet. The
Hamiltonian describing these excitations, for example, in the
point $\mathbf{K}$ coincides with the free Dirac one
\cite{Gonzales:1993:NP,Semenoff:1984:PRL} \be
\label{Hamiltonian.Kpoint} H = \sum_{\sigma = \uparrow,
\downarrow} \int \frac{d^2 k}{(2 \pi)^2} \bar{\psi}_{1 \sigma}(t,
\mathbf{k})(\bar{\gamma}^1 k_x + \bar{\gamma}^2 k_y) \psi_{1
\sigma}(t, \mathbf{k}), \ee where the momentum $\mathbf{k}
=(k_x,k_y)$ is already given in a local coordinate system
associated with a chosen K point, $\bar{\psi}_{1 \sigma} = \psi_{1
\sigma}^{\dagger} \bar{\gamma}^0$ and $\bar{\gamma}^{0,1,2} =
(\sigma_3, i \sigma_2, - i \sigma_1)$. This representation of
$\bar{\gamma}$-matrices can be mapped in the representation
(\ref{a.gamma}) by using a unitary transformation with $U =
(1/\sqrt{2})(\hat{I} + i \sigma_3)$. Since the local coordinate
system for the point $\mathbf{K}^\prime$ is related to the system
associated with the point $\mathbf{K}$ by a parity transformation,
these two spinors can be again combined in one four-component
Dirac spinor $\Psi_{\sigma} = (\psi_{1 \sigma}, \psi_{2 \sigma})$.
The number of spin components $N$ has to be regarded as an
adjustable parameter and $N =2$ corresponds to  the physical case.
Finally, the Lagrangian density of noninteracting quasiparticles
reads
\begin{equation}
\label{Lagrangian.Carbon}
\mathcal{L}_0 = \sum_{\sigma= 1}^{N} v_F \bar{\Psi}_{\sigma} (t, \mathbf{r}) \left(
\frac{i \gamma^0 (\partial_t + i \mu)}{v_F} -
i \gamma^1 \partial_x - i \gamma^2 \partial_y
\right) \Psi_{\sigma}(t, \mathbf{r}),
\end{equation}
where $\bar{\Psi}_\sigma = \Psi_{\sigma}^{\dagger} \gamma^0$ and
$4\times 4$ $\gamma$-matrices are either $(\sigma_3, i \sigma_3, -
i\sigma_1) \otimes \sigma_3$ \cite{Khveshchenko:2001:PRL} or can
be taken from the unitary equivalent representation
(\ref{gamma.matrices}) \cite{Semenoff:1984:PRL,Gorbar:2002:PRB}.
Since the terms with $\partial_{x,y}$ in
Eq.~(\ref{Lagrangian.Carbon}) originate from the usual kinetic
term of the tight-binding Hamiltonian, vector potential
$\mathbf{A}$ can be inserted in the Lagrangian
(\ref{Lagrangian.Carbon}) using a minimal coupling prescription.
Finally we note that three-dimensional version of the Dirac
Hamiltonian (\ref{Hamiltonian.Kpoint}) was used in
Ref.~\cite{Abrikosov:1998:PRB} to describe an unusual
magnetoresistance in the two doped chalcogenides Ag$_{2+\delta}$Se
and Ag$_{2+\delta}$Te.

\subsection{Model relativistic Lagrangian}
\label{sec:Dirac}

As we have seen, many models of planar condensed matter systems result
in the Dirac type form of the effective low energy theory. Thus as
a starting point of the present paper we choose the following Lagrangian
\begin{equation}
\label{Dirac.Lagrangian}
\mathcal{L} = \bar{\psi}_i (i \gamma^\mu D_\mu - \Delta) \psi_i,
\quad \mu = 0,1,2,
\end{equation}
where $D_{\mu} = \partial_\mu - i e A_\mu$ is the covariant derivative,
$\bar{\psi}_i \equiv \psi_i^{\dagger} \gamma^0$ is the Dirac conjugated
spinor and the vector potential for the external magnetic field $\mathbf{B}$
perpendicular to the plane is taken in the symmetric  gauge
\begin{equation}
\label{gauge}
\mathbf{A} = \left(-\frac{B}{2}x_2 , \frac{B}{2} x_1 \right).
\end{equation}
The nonzero chemical potential $\mu$ will also be taken into
account by choosing $A_0 = \mu/e$, in the energy-momentum space
this corresponds to a shifting $\omega \to \omega + \mu$. We
assume that the fermions carry an additional flavor index $i =1,
\ldots, N$ which can be used to calculate the sum over equivalent
nodes in the case of $d$SC case and to sum over the spin
components in the cases of DDW and graphite. In
(\ref{Dirac.Lagrangian}) we have already set the velocities $c =
v_{F} = v_{D} = v_{\Delta} = 1$, so that we consider the
``isotropic Dirac cone''. When it is needed they can be restored
according to the prescription discussed in
Refs.~\cite{Sharapov:2003:PRB,Ferrer:2003:EPJB}. The Dirac
$\gamma$-matrices are taken in the reducible four-component
representation and they obey the Clifford (Dirac) algebra, $
\gamma^\mu \gamma^\nu + \gamma^\nu \gamma^\mu = 2 \hat{I}_4 g^{\mu
\nu} $. The explicit form of the $\gamma$-matrices is given in
Eq.~(\ref{gamma.matrices}) and the symmetry properties of the
Lagrangian with reducible four-dimensional representation are
discussed, for example, in
\cite{Jackiw:1981:PRD,Gorbar:2002:PRB,Gusynin:1995:PRD}. We have
also included a mass (gap) term $\Delta$ in the Lagrangian
(\ref{Dirac.Lagrangian}). The physical origin of this gap depends
on the underlying system we consider.  For example, it is well
known that an external magnetic field is a strong catalyst in
generating such a gap for Dirac fermions (the phenomenon of
magnetic catalysis) \cite{Gusynin:1995:PRD}.
Usually the opening of the
gap marks an important transition which occurs in the system. In
particular, in the case of pyrolytic graphite a poor screening of the
Coulomb interaction may lead to excitonic instability resulting in
the opening of the gap in the electronic spectrum and manifesting
itself through the onset of an insulating charge density wave (see
e.g. \cite{Khveshchenko:2001:PRL,Gorbar:2002:PRB}). There are,
however, many obstacles in experimental detection of such a gap,
so that in the context of the MO studies we consider in
Secs.~\ref{sec:gap} and \ref{sec:Omega.T}  a possibility
for its detection using dHvA and SdH effects.

A full theory of MO should also include the Zeeman interaction
term which leads to a spin factor in the LK formula. Here,
however, we neglect this term motivated by the fact that for the
relativistic spectrum the MO may become  observable for the
relatively low fields when the spin splitting is still small. If
necessary, the Zeeman term can be added explicitly both to the
original Hamiltonian (\ref{Hamiltonian.DDW.matrix}) and the
Lagrangian (\ref{Dirac.Lagrangian}).

\section{Spectral function of Dirac quasiparticles in an external field}
\label{sec:GF}

The Green's function  of Dirac fermions described by the Lagrangian
(\ref{Dirac.Lagrangian}) in an external field given by the vector
potential (\ref{gauge}) reads
\begin{equation}
\label{Schwinger.representation}
S(x-y) = \exp \left( i e \int \limits_y^x A_{\lambda}
d z^{\lambda} \right)
{\tilde S}(x-y),
\end{equation}
where ${\tilde S}$ is the translation invariant part ${\tilde S}$ of
the Green's function. Its derivation using the Schwinger proper-time method
and decomposition over Landau level poles has been discussed in many papers
(see e.g.
Refs.~\cite{Chodos:1990:PRD,Gusynin:1995:PRD,Sharapov:2003:PRB}),
so that here we begin with the spectral function
associated with the translationary invariant part ${\tilde S}$ of the Green's
function
\begin{equation}
\label{spectral.def}
A(\omega, \mathbf{p}) = \frac{1}{2 \pi i} \left[
{\tilde S}^A(\omega - i0, \mathbf{p}) - {\tilde S}^R(\omega + i0, \mathbf{p}) \right],
\end{equation}
where the retarded, ${\tilde S}^R$, and advanced, ${\tilde S}^A$,
Greens' functions are written in the energy-momentum representation.
This spectral function decomposed over Landau levels reads
\cite{Ferrer:2003:EPJB}
\begin{equation}
\label{AD.clean}
A(\omega, {\bf p}) = \exp\left(-
\frac{{\bf p}^2}{|e B|}\right) \sum_{n=0}^{\infty} (-1)^n \left[
\frac{(\gamma^0 M_n + \Delta) f_1({\bf p}) + f_2({\bf p})}{2 M_n}
\delta(\omega - M_n) + \frac{(\gamma^0 M_n - \Delta) f_1({\bf p})
- f_2({\bf p})}{2 M_n} \delta(\omega + M_n) \right],
\end{equation}
where $M_n = \sqrt{\Delta^2 + 2 e B n}$ and
\begin{equation}
\label{f} f_1({\bf p}) = 2\left[ P_{-} L_n \left(2\frac{{\bf
p}^2}{|eB|}\right) - P_{+} L_{n-1}\left(2 \frac{{\bf p}^2}{|eB|}
\right) \right], \qquad f_2({\bf p}) = 4 (p_1 \gamma^1 + p_2
\gamma^2) L^{1}_{n-1}\left(2 \frac{{\bf p}^2}{|eB|} \right),
 \end{equation}
with $P_{\pm} = (1\pm  \mbox{sgn} (e B)i \gamma^1 \gamma^2)/2$
being projectors and $L_n$, $L_n^1$ Laguerre's polynomials
($L_{-1}^1 \equiv 0$). In what follows, for convenience, we take
$e B > 0$. The chemical potential $\mu$, as mentioned above, has
to be taken into account via the shift $\omega  \to \omega + \mu$.

In contrast to the nonrelativistic Landau levels with the energies
$E_{n} = \frac{e \hbar B}{mc} (n + \frac{1}{2})$ (here $m$ is the
effective mass of the carriers), in the relativistic problem with
zero gap $\Delta$ the energies are $E_n = \sqrt{\frac{e \hbar v_F
v_2}{c} B 2 n}$, $n =0,1, \ldots$, where we restored all
parameters ($v_{2} \equiv v_{F}, v_{D}$ or $v_{\Delta}$ depending
on the model we consider) to show explicitly the differences. For
the nonrelativistic problem the distance between Landau levels
coincides with the cyclotron frequency, $\omega_{c} = e \hbar B/(m
c)$, so that $\omega_c [\mbox{K}]\sim 1.35 B[\mbox{Tesla}] m_e/m$,
where $m_e$ is the electron mass, while for the relativistic
problem the corresponding energy scale, characterizing the
distance between Landau levels, is
\begin{equation}
\label{units}
\omega_L  = \sqrt{\frac{\hbar v_F v_2 2 e
B}{c}} [\mbox{K}] = 4.206 \times 10^{-4}
\sqrt{\frac{v_{2}}{v_{F}}} v_{F} [\mbox{m}/\mbox{s}] \sqrt{B
[\mbox{Tesla}]} ,
\end{equation}
where $v_{F}$ is given in $\mbox{m}/\mbox{s}$. For example,
choosing $m_e/m = 1$ we estimate that $\omega_c \sim 1 \mbox{K}
\cdot B[\mbox{Tesla}]$, but this value can be increased by using
metals such as Bi with a large ratio $m_e/m$. Making the estimate
for the relativistic case with a rather small $v_{F} = 2\times
10^5 \mbox{m/s}$ and $v_F/v_2 = 20$, which are typical values of
the parameters for the DDW model, we obtain $ \omega_L \sim 18K
\cdot \sqrt{B[\mbox{Tesla}]}$ which shows that in the systems with
the linear dispersion law the quantum condition favorable for MO
persists to rather high temperatures and small fields
\cite{Abrikosov:1998:PRB}. Moreover, repeating the estimate for
the Fermi velocity in graphite, $v_{F} = 9.7 \times 10^5
\mbox{m/s}$, we obtain even larger $ \omega_L \sim 400K \cdot
\sqrt{B[\mbox{Tesla}]}$. We note that since the Zeeman term has
the same magnitude as $\omega_c$, it is indeed small comparing to
$\omega_L$ and can be safely neglected.

In order  to consider the MO for a more realistic case, one should
introduce the effect of quasiparticle scattering that results in a
{\em Dingle\/} factor in the expression for the amplitude of MO.
In general, this can be done by considering dressed fermion
propagators that include the  self-energy $\Sigma(\omega)$ due to
the scattering from impurities. Up to now the problem of
scattering from impurities in the presence of a magnetic field
does not have yet a satisfactory solution. Therefore, here we
choose the case of constant width $\Gamma = \Gamma(\omega = 0) =
-\mbox{Im} \Sigma^R(\omega =0) = 1/(2 \tau)$, $\tau$ being a mean
free time of quasiparticles, so that the $\delta$-like
quasiparticle peaks corresponding to the Landau levels in
Eq.~(\ref{spectral.def}) acquire a Lorentzian shape:
\begin{equation}
\label{Gamma}
\delta(\omega \pm M_n) \to \frac{1}{\pi} \frac{\Gamma}{(\omega \pm M_n)^2 + \Gamma^2}.
\end{equation}

Such a broadening of Landau levels with a constant $\Gamma$ was
found to be a rather good approximation valid in not very strong
magnetic fields \cite{Prange.book,Shoenberg.book}. Definitely, the
treatment of disorder in the presence of the magnetic field in
such a simplified manner should be considered as only the first
step until further progress in this problem is achieved (in
connection with this, see, Ref.~\cite{Champel:2002:PRB}). The
technical advantage of the approximation $\Gamma=const$ is that
one can firstly consider dHvA oscillations with $\delta$-like
Landau levels at $T =0$ and only afterwards introduce the effect
of level broadening due to the finite temperature and
quasiparticle scattering convoluting the final results with the
appropriate distribution functions (see,
Refs.~\cite{Shoenberg.book,Shoenberg:1984:JLTP} and
Sec.~\ref{sec:Omega.general}). However, this simplification is
only valid if all Landau levels have the same width.

\section{Oscillations of density of states}
\label{sec:DOS}

We begin with the calculation of the quasiparticle density of
states  (DOS) $D_0(\epsilon)$ in the absence of scattering
($\Gamma =0$), which is expressed in terms of the spectral
function (\ref{AD.clean}) as
\begin{equation}
\label{DOS.def}
D_0(\epsilon)=\int_{\rm B}\frac{d^2p}{(2\pi)^2}{\rm
tr}\left[\gamma_0 A(\epsilon,\mathbf{p})\right],
\end{equation}
where the domain of integration $B$ is chosen to preserve the
volume of the original Brillouin zone. We also assume that the summation
over spin states that are identical in the absence of Zeeman
term is included in the definition of trace in $D(\epsilon)$. This, as
mentioned above, can be done by taking an appropriate value of
$N$.

Evaluating the trace and expanding the limits of integration over
momenta to $\infty$, we get the DOS as the sum of delta functions
of Landau's level energies:
\begin{equation}
 \label{DOS.delta}
D_0(\epsilon)=\frac{NeB}{2\pi}\left[\delta(\epsilon-\Delta)+\delta
(\epsilon+\Delta)+2\sum\limits_{n=1}^\infty\left(\delta(\epsilon-M_n)
+\delta(\epsilon+M_n)\right)\right].
\end{equation}
The fact that the original Brillouin zone has a finite volume can be taken
into account later using the finite limits of integration if necessary
and/or by implying that the DOS $D(\epsilon)$
is multiplied by the factor $\theta(\Lambda- |\epsilon|)$,
where $\Lambda$ is the bandwidth.

The broadening of Landau levels due to the scattering is taken
into account according to the prescription given by
Eq.~(\ref{Gamma}). It is easy to see that this prescription
corresponds to the convolution of the density of states
$D_0(\epsilon)$ with the Lorentz distribution
$P_\Gamma(\omega)=\Gamma/ [\pi(\omega^2 + \Gamma^2)]$
\begin{equation}
\label{q-p.DOS}
D(\epsilon) = \int \limits_{-\infty}^{\infty}
d \omega P_{\Gamma} (\omega  - \epsilon)D_0 (\omega),\quad
\int \limits_{-\infty}^{\infty} d \omega P_{\Gamma}(\omega) =1.
\end{equation}
In fact, instead of the Lorentz distribution any other normalized
probability  distribution can be used \cite{Shoenberg.book} in
order to introduce phenomenologically the energy level width. In
what follows we shall use the Lorentz distribution  which allows
analytical calculations. Physically, the Lorentz distribution
corresponds to an extremely strong disorder since all moments of
this distribution diverge. The quasiparticle DOS at the Fermi
surface can be obtained by evaluating Eq.~(\ref{q-p.DOS}) for
$\epsilon = \mu$. Eq.~(\ref{DOS.delta}) can be also rewritten as
\begin{equation}
\begin{split}
D_0(\epsilon)&=\frac{NeB|\epsilon|}{\pi}\left[\delta(\epsilon^2-\Delta^2)
+2\sum\limits_{n=1}^\infty\delta(\epsilon^2-\Delta^2-2eBn)\right]\\
&=\frac{NeB}{2\pi}{\rm
sgn}(\epsilon)\frac{d}{d\epsilon}\left[\theta(\epsilon^2-\Delta^2)
+2\sum\limits_{n=1}^\infty\theta(\epsilon^2-\Delta^2-2eBn)\right],
\label{DOS.theta}
\end{split}
\end{equation}
where $\theta$ is the step function.
Note that $NeB/(2\pi)$ is the density of the Landau levels.
Using the Poisson summation formula
\begin{equation}
\frac{1}{2}F(0)+\sum\limits_{n=1}^\infty F(n)=\int\limits_0^\infty
F(x)dx+2{\rm Re}\sum\limits_{k=1}^\infty \int\limits_0^\infty
F(x)e^{2\pi ikx}dx,
\end{equation}
we find the sum over the Landau levels
\begin{equation}
\sum\limits_{n=1}^\infty\theta\left(\epsilon^2-\Delta^2-2eBn\right)
=
\theta(\epsilon^2-\Delta^2)\left[-\frac{1}{2}+\frac{\epsilon^2-\Delta^2}{2eB}
+\sum\limits_{k=1}^\infty\frac{\sin\left( \pi
k\frac{\epsilon^2-\Delta^2}{eB}\right)}{\pi k}\right].
\end{equation}
Substituting now the last expression in Eq.(\ref{DOS.theta}) we
obtain the final expression for the DOS with zero broadening that
can be written in three equivalent forms:
\begin{subequations}
\label{DOS.oscillations}
\begin{align}
D_0(\epsilon)& = \frac{N}{2\pi}{\rm sgn}(\epsilon)
\frac{d}{d\epsilon}\left\{ \theta(\epsilon^2-\Delta^2)
\left[ \epsilon^2-\Delta^2
+2eB \sum\limits_{k=1}^\infty \frac{1}{\pi k} \sin \left(
\frac{\pi k (\epsilon^2-\Delta^2)}{eB} \right) \right] \right\}
\label{DOS.osc-first} \\
& =\frac{N}{2\pi}{\rm sgn}(\epsilon)\frac{d}{d\epsilon}
\left\{\theta(\epsilon^2-\Delta^2) \left[ \epsilon^2-\Delta^2
+\frac{2eB}{\pi}\tan^{-1} \left(\cot \left(
\frac{\pi(\epsilon^2-\Delta^2 )}{2eB} \right)
\right) \right] \right\}  \label{DOS.osc-second}\\
&=\frac{N}{2\pi}{\rm sgn}(\epsilon)\frac{d}{d\epsilon}
\left\{\theta(\epsilon^2-\Delta^2)\left[\epsilon^2-\Delta^2
-2eB\,B_1\left(\frac{\epsilon^2
-\Delta^2}{2eB}\right)\right]\right\} \label{DOS.osc-third}.
\end{align}
\end{subequations}
The summation in Eq.~\pref{DOS.osc-second} was done using the
formula \be \sum\limits_{n=1}^\infty\frac{\sin(\pi n x)}{n}=
\tan^{-1}\left( \frac{\sin(\pi x)}{1-\cos(\pi x)} \right). \ee and
Eq.~\pref{DOS.osc-third} is written using the Bernoulli
polynomials $B_n(x)$ periodically continued beyond the interval
$[0,1]$: \be \label{Bernoulli.def} B_n(x)=-\frac{2\cdot
n!}{(2\pi)^n}\sum\limits_{k=1}^\infty\frac{1}{k^n} \cos\left(2\pi
kx-\frac{n\pi}{2}\right),\quad n\geq2,\quad 0\leq x\leq1; \quad
n=1,\quad 0<x<1. \ee In fact, all the Bernoulli polynomials we
dealt with in this paper depend on the fractional part $mod[x]$ of
their argument $x$, i.e., $B_n(mod[x])$ (here $mod[x]$ is a
shorthand notation for $x$ modulo $1$, i.e.,  $mod[x]=x-[x]$ where
$[x]$ is the largest integer satisfying $[x]\leq x$). In what
follows, for brevity we omit the sign $mod[]$ and write simply
$B_n(x)$.

\subsection{DOS in the presence of scattering}

Now we turn to calculating DOS in presence of the impurity
scattering rate $\Gamma$ using convolution (\ref{q-p.DOS}). Since
$D_0(\epsilon)$ is an even function of $\epsilon$, the expression
for $D(\epsilon)$ can be written as follows
\begin{subequations}
\label{DOS_through_convolution}
\begin{align}
& D(\epsilon)=\frac{N}{\pi^2}{\rm
Im}\left\{(\epsilon+i\Gamma)\left[
\int\limits_0^{\Lambda^2/eB}\frac{dx}{x+z} +\frac{2}{\pi}
\int\limits_0^{\infty} \frac{dx}{(x+z)^2}
\tan^{-1}\left(\cot\frac{\pi x}{2}\right)
\right]\right\}, \label{DOS.main}\\
& x = \frac{\omega^2 - \Delta^2}{eB}, \qquad
z=\frac{\Delta^2-(\epsilon+i\Gamma)^2}{eB},
\end{align}
\end{subequations}
where we used Eq.~\pref{DOS.osc-second} for $D_0(\epsilon)$,
represented $P_{\Gamma}$ as $P_{\Gamma}(\epsilon \pm \omega ) = -
\mbox{Im} (\epsilon \pm \omega + i \Gamma)^{-1}$
and integrated by parts to obtain the second term in square brackets in
Eq.~(\ref{DOS.main}). We also put an
explicit cutoff $\Lambda$ associated with the bandwidth in the
first integral describing non-oscillating part of DOS.

It is easy to calculate that
\be
{\rm Im}\left[(\epsilon+i\Gamma)
\int\limits_0^{\Lambda^2/eB}\frac{dx}{x+z}\right]
=\Gamma\ln\frac{\Lambda^2}{\sqrt{(\epsilon^2-\Delta^2-\Gamma^2)^2+
4\epsilon^2\Gamma^2}}+\epsilon\tan^{-1}\frac{2\epsilon\Gamma}{\Delta^2
+\Gamma^2-\epsilon^2}.
\ee

The second integral in Eq.~\pref{DOS_through_convolution} denoted
as $I$ can be calculated if we take into account that the function
$\tan^{-1}\left(\cot({\pi x}/{2})\right)$ is periodic with the
period $x=2$ so that we can write \be
\begin{split}
I&=\frac{2}{\pi}\int\limits_0^{\infty}
\frac{dx}{(x+z)^2}\tan^{-1}\left(\cot\frac{\pi x}{2}\right)=\frac{2}{\pi}
\sum\limits_{k=0}^\infty\int\limits_{2k}^{2k+2}\frac{dx}{(x+z)^2}\tan^{-1}
\left(\cot\frac{\pi x}{2}\right)\\
&=\frac{2}{\pi}
\sum\limits_{k=0}^\infty\int\limits_{0}^{2}\frac{dx}{(x+2n+z)^2}\tan^{-1}
\left(\cot\frac{\pi x}{2}\right)=\frac{1}{2}\int\limits_0^1dx(1-2x)\zeta(2,
x+z/2),
\end{split}
\ee
where $\zeta(s,x)$ is the generalized zeta-function and we
used also that $\tan^{-1}\left(\cot({\pi
x}/{2})\right)=(\pi/2)(1-x)$ for $0 < x < 2$. The last integral
is evaluated using the properties of $\zeta$-function
\be
\int\limits_0^1dx\zeta(2,x+z)=\frac{1}{z},\quad
\int\limits_0^1dx\,x\zeta(2,x+z)=\Psi(1+z)-\ln z, \ee so that we
finally get \be \label{integral-I}
I=-\psi\left(\frac{z}{2}\right)+\ln\frac{z}{2}-\frac{1}{z}. \ee
Thus, we express the DOS in terms of the digamma function $\psi$:
\be
\label{DOS_through_psi}
\begin{split}
D(\epsilon)=\frac{N}{\pi^2}\left\{\Gamma\ln\frac{\Lambda^2}
{2eB}-{\rm Im}\left[(\epsilon+i\Gamma)\left(\psi
\left(\frac{\Delta^2-(\epsilon+i\Gamma)^2}{2eB}\right)+\frac{eB}{\Delta^2
-(\epsilon+i\Gamma)^2}\right)\right]\right\},
\end{split}
\ee
or, in equivalent form
\be
D(\epsilon)=\frac{N}{\pi^2}\Gamma\ln\frac{\Lambda^2}
{2eB}+\frac{NeB}{\pi^2}\frac{d}{d\epsilon}{\rm
Im}\left[\ln\Gamma\left(
\frac{\Delta^2-(\epsilon+i\Gamma)^2}{2eB}\right)+
\frac{1}{2}\ln\left(\frac{\Delta^2-(\epsilon+i\Gamma)^2}{2eB}\right)\right].
\label{DOS-through-logGamma}
\ee
It is evident that the DOS oscillations are contained in the $\psi$-function
when the real part of its argument becomes negative.
They can be extracted
in explicit form using the relationship
\be
\psi(-z)=\psi(z)+\frac{1}{z}+\pi\cot(\pi z).
\ee
Hence we get
\be
\begin{split}
&\psi\left(\frac{\Delta^2-(\epsilon+i\Gamma)^2}{2eB}\right)=
\psi\left(\frac{\Delta^2-(\epsilon+i\Gamma)^2}{2eB}\right)
\left(\theta(\epsilon^2-\Delta^2-\Gamma^2)+\theta(\Delta^2+\Gamma^2-
\epsilon^2)\right)\\
&={\rm Re}\psi\left(\frac{|\epsilon^2-\Delta^2-\Gamma^2|-2i\epsilon\Gamma}
{2eB}\right)-i\,{\rm sgn}(\epsilon^2-\Delta^2-\Gamma^2){\rm Im}\psi\left(
\frac{|\epsilon^2-\Delta^2-\Gamma^2|-2i\epsilon\Gamma}{2eB}\right)\\
&+\theta(\epsilon^2-\Delta^2-\Gamma^2)\left[\frac{2eB}{\epsilon^2
-\Delta^2-\Gamma^2+2i\epsilon\Gamma}+\pi\cot\pi\frac{\epsilon^2
-\Delta^2-\Gamma^2+2i\epsilon\Gamma}{2eB}\right],
\end{split}
\ee
so that Eq.~\pref{DOS_through_psi} is rewritten as
\be
\label{DOS.Re-Im}
\begin{split}
D(\epsilon)&=\frac{N}{\pi^2}\left\{\Gamma\left[\ln\frac{\Lambda^2
}{2eB}-{\rm Re}\psi\left(\frac{|\epsilon^2-\Delta^2-\Gamma^2|-
2i\epsilon\Gamma}{2eB}\right)-\frac{eB|\epsilon^2-\Delta^2-\Gamma^2|}
{(\epsilon^2-\Delta^2-\Gamma^2)^2+4\epsilon^2\Gamma^2}\right]\right.\\
&+\left.\epsilon
{\rm sgn}(\epsilon^2-\Delta^2-\Gamma^2)\left[{\rm Im}
\psi\left(\frac{|\epsilon^2-\Delta^2-\Gamma^2|-2i\epsilon\Gamma}{2eB}\right)
+\frac{2eB\epsilon\Gamma}
{(\epsilon^2-\Delta^2-\Gamma^2)^2+4\epsilon^2\Gamma^2}
\right]\right.\\
&+\left.\pi\theta(\epsilon^2-\Delta^2-\Gamma^2)\frac{\epsilon\sinh(2\pi
\epsilon\Gamma/eB)-\Gamma
\sin[\pi(\epsilon^2-\Delta^2-\Gamma^2)/eB]}{\cosh(2\pi\epsilon\Gamma/eB)
-\cos[\pi(\epsilon^2-\Delta^2-\Gamma^2)/eB]}\right\}.
\end{split}
\ee
For small $\Gamma$ the main contribution comes from the last
term in Eq.(\ref{DOS.Re-Im}), thus DOS in the presence of $\Gamma$
can be represented in the form similar to Eq.~(\ref{DOS.osc-first}):
\be
\begin{split}
\label{dos-oscil-gamma}
D(\epsilon)&=\frac{N}{2\pi}
{\rm sgn}(\epsilon)\frac{d} {d\epsilon}\left\{\theta(\epsilon^2
-\Delta^2-\Gamma^2)\left[\epsilon^2-\Delta^2-\Gamma^2\right.\right. \\
&+\left.\left.2eB\sum
\limits_{k=1}^\infty\frac{1}{\pi k}\sin\frac{\pi k(\epsilon^2-
\Delta^2-\Gamma^2)}{eB}\exp\left(-\frac{2\pi k|\epsilon|\Gamma}
{eB}\right)\right]\right\}.
\end{split}
\ee
Using now the formula
\be
\label{sumformula1}
\sum\limits_{k=1}^\infty\frac{1}{k}e^{-kt}\sin kx=
\tan^{-1}\frac{\sin x}{e^t-\cos x}, \quad t>0,
\ee
one can check that the last term in Eq.~(\ref{DOS.Re-Im}) is recovered.
It is evident that oscillating part of DOS is contained in the sum over $k$ in
Eq.~(\ref{dos-oscil-gamma}).

\subsection{DOS in limiting cases}

Since the final expressions for DOS, Eq.(\ref{DOS_through_psi})
and especially Eq.(\ref{DOS.Re-Im}) are rather lengthy, it is
useful to consider a few simple limiting cases and compare them
with the known results. First of all, it is easy to obtain from
Eq.~(\ref{DOS_through_psi}) that in the limit of zero field $B=0$
the DOS becomes \be \label{E.B=0}
D(\epsilon) =  \frac{N}{\pi^2} \left[ \Gamma \ln
\frac{\Lambda}{\sqrt{\Gamma^2 + (\epsilon- \Delta)^2}} + \Gamma
\ln \frac{\Lambda}{\sqrt{\Gamma^2 + (\epsilon+ \Delta)^2}} +
|\epsilon| \left(
\frac{\pi}{2}+\tan^{-1}\frac{\epsilon^2-\Delta^2-\Gamma^2}{2|\epsilon|\Gamma}
\right) \right],
\ee
and for $\Delta = 0$, after restoring the prefactor $1/(v_F
v_\Delta)$ with the velocities $v_F$ and $v_{\Delta}$, it reduces
to the DOS derived in Ref.~\cite{Durst:2000:PRB}.

As follows from Eq.~(\ref{DOS.Re-Im}), the DOS at zero energy, but
in the finite field is given by \be \label{E=0.B}
D(0)=\frac{N\Gamma}{\pi^2}\left[\ln\frac{\Lambda^2}{\Delta^2
+\Gamma^2}-\Psi\left(\frac{\Delta^2+\Gamma^2}
{2eB}\right)+\ln\frac{\Delta^2+\Gamma^2}{2eB}-\frac{eB}{\Delta^2+\Gamma^2}
\right]. \ee The first term of Eq.~(\ref{E=0.B}) is nothing else
but the zero energy DOS (\ref{E.B=0}) in the absence of the
magnetic field. The behavior of the DOS (\ref{E=0.B}) can be now
studied in various asymptotical regimes. For example, for
$\Delta=0$ and $\Gamma\to0$ we find \be \label{E=0.B.Gamma-0}
D(0)=\frac{N}{\pi^2}\left[\frac{eB}{\Gamma}+\Gamma\ln\frac{\Lambda^2}
{2eB}\right], \ee i.e., DOS is enhanced in the presence of the
magnetic field if $\Gamma^2\ll eB$. In the opposite limit, $eB\ll
\Gamma^2$ we have \be
D(0)=\frac{N\Gamma}{\pi^2}\left[\ln\frac{\Lambda^2}
{\Gamma^2}+\frac{(eB)^2}{3\Gamma^4}\right]. \ee On the other hand,
for $\Delta\neq0$ and $\Gamma\to0$ we obtain \be
D(0)=\frac{N\Gamma}{\pi^2}\left[\ln\frac{\Lambda^2}
{2eB}-\Psi\left(\frac{\Delta^2}
{2eB}\right)-\frac{eB}{\Delta^2}\right], \ee so that nonzero gap
regularizes $\Gamma \to 0$ divergence, which is present in
Eq.~(\ref{E=0.B.Gamma-0}).

\section{Representation for thermodynamic potential}
\label{sec:Omega.general}

\subsection{General expressions}

All thermodynamic quantities such as the magnetization
$\mathbf{M}$ and the number of electrons $N$ can be found from the
grand thermodynamic potential $\Omega$, which in the relativistic
case (see Refs.~\cite{Gusynin:1995:PRD,Cangemi:1996:AP}) can be
written using the DOS as follows \be \label{Omega.def}
\Omega(T,\mu)=-T\int\limits_{-\infty}^\infty d\epsilon
D(\epsilon)\ln\left( 2\cosh\frac{\epsilon-\mu}{2T}\right) \ee with
the DOS $D(\epsilon)$ given by Eq. (4.3). We assume everywhere
that the volume (area) of the system is unity, so that
Eq.~(\ref{Omega.def}) corresponds to the thermodynamic potential
per unit volume. It is convenient to separate the vacuum
contribution at $T=0$, $\mu=0$ in the thermodynamic potential. For
that we write \be
\begin{split}
\Omega(T,\mu)&=-T\int\limits_{-\infty}^\infty d\epsilon\, D(\epsilon)
\left[\ln\left(e^{\frac{\epsilon-\mu}{2T}}+
e^{\frac{\mu-\epsilon}{2T}}\right)
\left(\theta(\epsilon)+\theta(-\epsilon)\right)\right]\\
&=-\frac{1}{2}\int\limits_{-\infty}^\infty d\epsilon\, D(\epsilon)
(\epsilon-\mu){\rm sgn}(\epsilon)-T\int\limits_{-\infty}^\infty d\epsilon\,
D(\epsilon)\left[\ln\left(1+e^{\frac{\mu-\epsilon}{T}}\right)\theta(\epsilon)
+\ln\left(1+e^{\frac{\epsilon-\mu}{T}}\right)\theta(-\epsilon)\right],
\end{split}
\ee or, using the evenness of the function
$D(\epsilon)$:
\be \label{Omega.def.final}
\Omega(T,\mu)=-\int\limits_{0}^\infty d\epsilon\, \epsilon
D(\epsilon)-T\int \limits_{-\infty}^\infty d\epsilon\,
D(\epsilon)\left[\ln\left(1+e^{\frac{\mu-\epsilon}{T}}\right)\theta(\epsilon)
+\ln\left(1+e^{\frac{\epsilon-\mu}{T}}\right)\theta(-\epsilon)\right].
\ee The first (divergent) term in the last expression is the
vacuum energy while the second one (convergent) is due to
contributions of real quasiparticle excitations.

At zero temperature, $T=0$, we thus have
\be
\label{vacuum+}
\Omega(0,\mu)= \Omega(0, \mu=0) +\int\limits_0^{|\mu|}d\epsilon\, D(\epsilon)(\epsilon
-|\mu|)\equiv \Omega_0(0)+\tilde{\Omega}_0(\mu).
\ee
It is easy to see that the density of states is related to the thermodynamic
potential at zero temperature
\be
D(\mu)=-\frac{d^2\Omega(T=0,\mu)}{d\mu^2}.
\ee

As was already mentioned, the constant Landau level width
approximation appears to be very useful simplification, viz.
instead of calculating the potential (\ref{Omega.def}) with
$\Gamma \neq 0$, one can start from the potential $\Omega_T$, \be
\Omega_T(\mu)=-T\int\limits_{-\infty}^\infty d\epsilon
D_0(\epsilon)\ln\left( 2\cosh\frac{\epsilon-\mu}{2T}\right),
\label{OmegaT.def} \ee where $D_0(\epsilon)$ is the DOS with zero
level broadening ($\Gamma =0$). Then the potential $\Omega$ at
finite $\Gamma$ is obtained by convoluting $\Omega_T(\mu)$ with
the distribution function $P_\Gamma(\omega)$. The statement that
the level broadening is equivalent to considering the distribution
of the chemical potentials was proven in
Ref.~\cite{Shoenberg:1984:JLTP} and we adopt here its derivation
for the relativistic thermodynamic potential Eq.(\ref{Omega.def}).
Indeed, if the levels are broadened, then the density of states is
obtained as a convolution of $D_0(\epsilon)$ with the probability
distribution $P_\Gamma(\epsilon)$ of energies $\epsilon$: \be
\label{Omega.PG}
\begin{split}
\Omega(T,\mu) & = -T\int\limits_{-\infty}^\infty d\omega d\epsilon\,
P_\Gamma(\omega-\epsilon)\,D_0(\omega)\ln\left(2\cosh\frac{\epsilon-\mu}{2T}
\right)=-T\int\limits_{-\infty}^\infty d\omega d\epsilon\,
P_\Gamma(\omega)D_0(\epsilon)\ln\left(2\cosh\frac{\epsilon-\omega-\mu}{2T}
\right)\\
&=-T\int\limits_{-\infty}^\infty d\omega d\epsilon\,
P_\Gamma(\omega-\mu)\,D_0(\epsilon)\ln\left(2\cosh\frac{\epsilon-\omega}{2T}
\right)=\int\limits_{-\infty}^\infty d\omega\, P_\Gamma(\omega-\mu)
\Omega_T(\omega),
\end{split}
\ee
where the potential $\Omega_T(\mu)$ is given by Eq.(\ref{OmegaT.def}).
If several damping effects occur together, the corresponding convolutions
have to be carried out successively (the order is not essential).

The effect of finite temperature can also be included in this
scheme by choosing the distribution function $P_T(z)$, which
describes the temperature line broadening, equal to the negative
derivative of the Fermi function, i.e.
\begin{equation}
P_T(z)=-\frac{\partial n_F(z)}{\partial z}=\frac{1}{4T\cosh^2 \frac{z}{2T}}.
\label{therm-distribution}
\end{equation}
We now show that the thermodynamic potential
Eq.~(\ref{OmegaT.def}) can be obtained as the convolution of
$\Omega_0(\mu)$ with the distribution function $P_T$ given by
Eq.~(\ref{therm-distribution})
\begin{equation}
\Omega_T(\mu)=\int\limits_{-\infty}^\infty d\omega P_T(\omega-\mu)
\Omega_0(\omega).
\label{thermal_convolution}
\end{equation}
Since the vacuum contribution does not change under averaging over
thermal and $\Gamma$ distributions, it is enough to prove this
fact for the finite part $\tilde{\Omega}_T(\mu)$. Performing
integration by parts we obtain:
\be \label{Omega.T}
\tilde{\Omega}_T(\mu)=-T\int \limits_{-\infty}^\infty d\epsilon\,
D(\epsilon)\left[\ln\left(1+e^{\frac{\mu-\epsilon}{T}}\right)\theta(\epsilon)
+\ln\left(1+e^{\frac{\epsilon-\mu}{T}}\right)\theta(-\epsilon)\right]=
-\int\limits_{-\infty}^\infty d\epsilon f_F(\epsilon)
\int\limits_0^\epsilon dx\,{\rm sgn}(x) D(x), \ee where
$f_F(\epsilon)$ is the relativistic thermal distribution function
\be
f_F(\epsilon)=\frac{\theta(\epsilon)}{1+e^{\frac{\epsilon-\mu}{T}}}+
\frac{\theta(-\epsilon)}{1+e^{\frac{\mu-\epsilon}{T}}}, \ee which
is also used in semiconductor physics. Integrating by parts for
the second time, we obtain \be
\tilde{\Omega}_T(\mu)=\int\limits_{-\infty}^\infty \frac{\partial
n_F(\epsilon-\mu)}{\partial\epsilon}W(\epsilon)d\epsilon, \ee
where \be W(\epsilon)=\int\limits_0^\epsilon dx\,{\rm
sgn}(x)\int\limits_0^x dy\,
 {\rm sgn}(y)D(y).
\ee Since \be -\frac{\partial
n_F(\epsilon)}{\partial\epsilon}=\frac{1} {4T\cosh^2
\frac{\epsilon}{2T}}\rightarrow \delta(\epsilon), \quad T\to0, \ee
we find that \be W(\mu)=-\tilde{\Omega}_0(\mu), \ee hence \be
\tilde{\Omega}_T(\mu)=\int\limits_{-\infty}^\infty d
\epsilon\left(-\frac{\partial
n_F(\epsilon-\mu)}{\partial\epsilon}\right)
\tilde{\Omega}_0(\epsilon). \ee The common effect of level
broadening due to both temperature and damping effects can be
written as successive convolutions
\begin{equation}
\label{Omega.PG-PT}
\Omega(T,\mu)=\int\limits_{-\infty}^{\infty}d\omega^\prime d\omega
P_\Gamma(\omega^\prime-\mu)P_T(\omega-\omega^\prime)\Omega_0(\omega).
\end{equation}
Hence for calculating thermodynamic quantities we need to know
only  the thermodynamic potential $\Omega_0(\omega)$ at zero
temperature and zero width of levels that is directly expressed
via the DOS $D_0(\epsilon)$, Eq.~(\ref{DOS.oscillations}).
Thus the knowledge of zero $T$
(and zero $\Gamma$ if the Landau levels have the same width) DOS is completely
sufficient to write down the finite $T$ thermodynamic potential.

\subsection{Link with nonrelativistic thermodynamic potential and
equation for chemical potential}
\label{sec:number.eq}

We start from the expression for the nonrelativistic thermodynamic potential
(see, e.g. Ref.~\cite{Shoenberg.book})
\be
\label{Omega.nonrel.def}
\Omega_{\rm NR}(T,\mu) = - T \int\limits_{-\infty}^{\infty} d \epsilon
\mathcal{D}(\epsilon) \ln \left(1 + e^{\frac{\mu - \epsilon}{T}} \right).
\ee
expressed in terms of the DOS $\mathcal{D} (\epsilon)$
for original nonrelativistic tight-binding Hamiltonians considered
in Sec.~\ref{sec:model}. Due to the complexity of the tight-binding spectrum,
this DOS cannot be found explicitly for $B \neq 0$ and
the only property of the DOS we need is that it is an even function
of $\epsilon$.

The derivative of the thermodynamic potential
(\ref{Omega.nonrel.def}) with respect to the chemical potential
$\mu$, \be \label{number.nonrelativistic}
-\frac{\partial\Omega_{\rm NR}(T,\mu)}{\partial\mu}=
\int\limits_{-\infty}^\infty d\epsilon
\mathcal{D}(\epsilon)n_F(\epsilon-\mu) =n, \ee determines the
density of carriers $n$ in nonrelativistic many-body theory as a
function of $T$, $B$ and $\mu$. On the other hand one can consider
Eq.~(\ref{number.nonrelativistic}) as an equation for $\mu$ as a
function of $T$, $B$ and $n$ which is typical for studies in a
canonical ensemble. In Eq.~(\ref{number.nonrelativistic})
$n_{F}(\omega) = 1/[\exp(\omega/T) +1]$ is the usual Fermi
function. At $T=0,\mu=0$ Eq.~(\ref{number.nonrelativistic}) gives
the density of particles for a half-filled band. \be
n_0=\int\limits_{-\infty}^\infty d\epsilon
\mathcal{D}(\epsilon)\theta (-\epsilon), \ee so that the deviation
of the particle density from $n_0$ is due to a finite temperature
and nonzero chemical potential.

It is convenient to redefine the thermodynamic potential in order
to have an explicit proportionality of $\mu$ to the density of
free carriers. Thus we introduce \be
\Omega^\prime(T,\mu)=\Omega_{\rm NR}(T,\mu)+\mu
n_0=-\int\limits_{-\infty}^\infty d\epsilon \mathcal{D}(\epsilon)
\left[T\ln \left(1 + e^{\frac{\mu - \epsilon}{T}} \right)
-\mu\theta(-\epsilon)\right]. \ee Inserting
$1=\theta(\epsilon)+\theta(-\epsilon)$ before the logarithm, the
last expression is rewritten in the form \be \label{Omega.prime}
\Omega^\prime(T,\mu)=-\int\limits_{0}^\infty d\epsilon\, \epsilon
\mathcal{D}(\epsilon)-T\int \limits_{-\infty}^\infty d\epsilon\,
\mathcal{D}(\epsilon)\left[\ln\left(1+e^{\frac{\mu-\epsilon}{T}}\right)\theta(\epsilon)
+\ln\left(1+e^{\frac{\epsilon-\mu}{T}}\right)\theta(-\epsilon)\right],
\ee where the first term gives the energy due to the half-filled
zone. As is seen, $\Omega^\prime(T,\mu)$ acquires the form of a
relativistic thermodynamic potential (compare with
Eq.~(\ref{Omega.def.final})). Its derivative with respect to $\mu$
\be \label{imbalance}
\begin{split}
\rho&=-\frac{\partial\Omega^\prime(T,\mu)}{\partial\mu}= \int\limits_{-\infty}^{\infty}
d \epsilon \mathcal{D}(\epsilon) \left[
n_{F}(\epsilon -\mu) \theta(\epsilon) - [1 - n_{F}(\epsilon -\mu)] \theta(-\epsilon)
\right]\\
&=  -\frac{1}{2}\int_{-\infty}^{\infty} d \epsilon \mathcal{D}(\epsilon)
\tanh \frac{\epsilon-\mu}{2T},
\end{split}
\ee determines the charge density of carriers or carrier
imbalance, $\rho$ ($\rho \equiv n - n_0= n_+ - n_-$,  where $n_+$
and $n_-$ are the densities of ``electrons'' and ``holes,
respectively) as a function of $T$ and $B$ at fixed $\mu$. Note
that now $\rho=0$ at $\mu=0$ and there is a symmetry with respect
to the transformation $\mu \to - \mu$ and $\rho \to - \rho$. In
the present paper, however, we will mostly use the canonical
ensemble interpretation of Eq.~(\ref{imbalance}), viz. as the
equation for $\mu$ as a function of $T$ and $B$ at fixed $\rho$.

\section{Zero temperature thermodynamic potential}
\label{sec:Omega0.Gamma=0} The zero temperature thermodynamic
potential is given by  Eq.(\ref{vacuum+}), and for calculation of
its vacuum contribution $\Omega_0(0)$ we refer to
Appendix~\ref{sec:A}. The thermodynamic potential with vacuum
energy $\Omega_{0}(0)$ subtracted is given  by the second term of
Eq.~(\ref{vacuum+}) \be \label{omega-subtracted.Gamma=0}
\tilde{\Omega}_0(\mu)=\int\limits_0^{|\mu|}d\epsilon
D_0(\epsilon)(\epsilon -|\mu|), \ee where the DOS for $\Gamma = 0$
case is given by Eq.~(\ref{DOS.oscillations}). In what follows we
choose for definiteness $\mu>0$. Calculating
$\tilde{\Omega}_0(\mu)$ (see Appendix~\ref{sec:B} for the detail)
we obtain that it can be represented as a sum of regular and
oscillating terms, \be \label{Omega.reg+osc}
\tilde{\Omega}_0(\mu)=\Omega_{reg}(\mu) + \Omega_{osc}(\mu), \ee
with $\Omega_{reg}(\mu)$ expressed in terms of generalized zeta
function \be \label{Omega.T=0.reg}
\Omega_{reg}(\mu)=-\frac{N}{2\pi}\theta(\mu-\Delta)\left[\frac{1}{3}
\mu(\mu^2-3\Delta^2)-\Delta eB
-(2eB)^{3/2}\zeta\left(-\frac{1}{2},
1+\frac{\Delta^2}{2eB}\right)\right], \ee and \be
\label{Omega.T=0.osc} \Omega_{osc}(\mu)=\frac{N(eB)^{3/2}}{2 \pi}
\theta (\mu - \Delta) \sum\limits_{k=1}^\infty \frac{1}{(\pi
k)^{3/2}}\left[ J_1(\pi kv)\cos(\pi k w)+J_2(\pi kv)\sin(\pi k
w)\right], \ee where the monotonic functions $J_{1,2}$ are defined
by Eq.~(\ref{J.def}). Writing the last expression for
$\Omega_{osc}(\mu)$ we introduced new variables \be \label{w-v}
w=\frac{\mu^2-\Delta^2}{eB},\quad v=\frac{\mu^2}{eB}. \ee

\begin{figure}[h]
\centering{
\includegraphics[width=8cm]{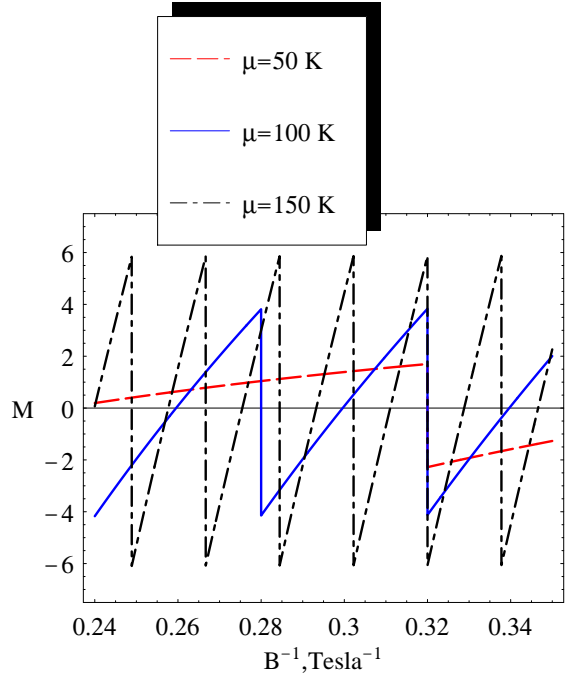}}
\caption{The magnetization $M$ (in $\mbox{K}/(\mbox{Tesla cm}^2)$
calculated for $N=1$) as a function of inverse field, $B^{-1}$,
for three different values of chemical potential and $\Delta= T =
\Gamma=0$. We use $eB \to 200 \mbox{K}^2 \cdot B[\mbox{Tesla}]$.}
\label{fig:1}
\end{figure}

\begin{figure}[h]
\centering{
\includegraphics[width=8cm]{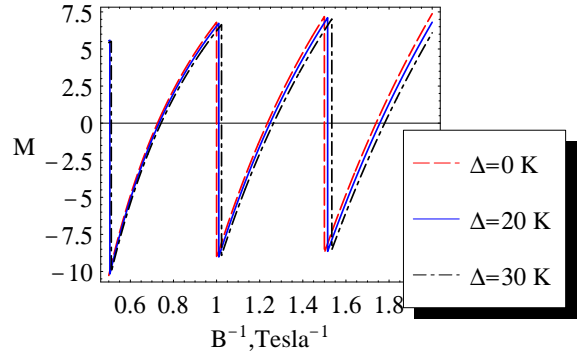}}
\caption{The magnetization $M$ (in $\mbox{K}/(\mbox{Tesla cm}^2)$
calculated for $N=1$) as a function of inverse field, $B^{-1}$,
for three different values of the gap $\Delta$ and for $T = \Gamma
=0$. We use $eB \to 10^4 \mbox{K}^2 \cdot B[\mbox{Tesla}]$ and
$\mu = 200 \mbox{K}$ that are typical for graphite.} \label{fig:2}
\end{figure}

For small fields, $eB<<\mu^2$, we can use the asymptotic expansion
(\ref{J_int_asym}) for $J_1,J_2$ and represent $\Omega_{osc}$ in
terms of periodically continued Bernoulli polynomials defined in
Eq.~(\ref{Bernoulli.def}): \be \label{Omega.T=0.osc.Bernoulli}
\Omega_{osc}(\mu)=\frac{N(eB)^2}{\pi^{3/2}\mu} \theta(\mu
-\Delta)\sum
\limits_{n=0}^\infty\frac{\Gamma(n+1/2)B_{n+2}(w/2)}{(n+2)!}
\left(\frac{2eB}{\mu^2}\right)^n. \ee In particular, keeping the
first few terms in the expansion we have \be
\label{Omega.T=0.osc.fewBernoulli}
\Omega_{osc}(\mu)\simeq\frac{N(eB)^2}{2\pi\mu} \theta(\mu -
\Delta) \left[B_2 \left(\frac{w}{2} \right) +\frac{eB}{3\mu^2}B_3
\left(\frac{w}{2} \right)+\frac{(eB)^2}{4\mu^4} B_4
\left(\frac{w}{2} \right)\right], \ee where the explicit
expressions for the Bernoulli polynomials $B_{1}$ -- $B_{4}$ are
\be \label{lowest.Bernoulli} B_1(x) = x -1/2, \quad
B_2(x)=x^2-x+1/6,\quad B_3(x)=x^3-3x^2/2+x/2,\quad B_4(x)
=x^4-2x^3+x^2-1/30. \ee The terms oscillating with a magnetic
field (Eqs.~(\ref{Omega.T=0.osc.Bernoulli}),
(\ref{Omega.T=0.osc.fewBernoulli})) represent small corrections to
the nonoscillating part of the thermodynamic potential
(Eq.(\ref{Omega.T=0.reg})) at $eB \ll \mu^2$. Nevertheless, since
$\Omega_{osc}$ contains fast oscillating functions of the variable
$(\mu^2-\Delta^2)/2eB$, after differentiating it over the magnetic
field,  the magnetization, $M=-\partial\Omega/\partial B$,
acquires a large oscillating part (see
Sec.~\ref{sec:magnetization} and Figs.~\ref{fig:1}, \ref{fig:2} ).

\subsection{The period of oscillations and its dependence on $\Delta$
for fixed $\mu$}
\label{sec:gap}

As it follows from the expressions (\ref{Omega.T=0.osc}),
(\ref{Omega.T=0.osc.Bernoulli}), the frequency of oscillations of
the thermodynamic potential is equal to $w/2 = (\mu^2 -
\Delta^2)/2eB$. Therefore their period $\delta B$ satisfies the
relationship \be \label{period} \frac{\delta B}{B} = \frac{2
eB}{\mu^2 - \Delta^2} \to
 \frac{\omega_L^2}{\mu^2 - \Delta^2},
\qquad \delta B \ll B,
\ee
where $\omega_L$ is defined in Eq.~(\ref{units}).
Using the estimate of $\omega_L$ for the DDW model made below
Eq.~(\ref{units}), we obtain that for $\Delta =0$,
$\mu = 100 K$ and $B = 1 \mbox{Tesla}$, the period
of oscillation $\delta B = 0.032 \mbox{Tesla}$, which is clearly within the
detectable range of dHvA oscillation experiments and close to the estimate
$\delta B \approx 0.05 \mbox{Tesla}$ made in Ref.~\cite{Nguyen:2002:PRB}.

Repeating this estimate with $B = 1 \mbox{Tesla}$ for graphite
with $\mu = 200 K$ and $\Delta =0$, we get $\delta B = 4
\mbox{Tesla}$ which is very close to the observed value for period
of the SdH oscillations in this material \cite{Wang:2003:PLA}.
However, so large value of $\delta B$ clearly violates the
condition $\delta B \ll B$ implied to obtain Eq.~(\ref{period}),
and for the case of graphite the period of MO is given by a more
accurate expression \be \label{period.graphite} \left|\frac{1}{e
B_{i}} - \frac{1}{e B_{i+1}} \right| = \frac{2}{\mu^2 - \Delta^2}
= \frac{2 \pi}{A}, \ee where $B_i$ and $B_{i+1}$ are the values of
the magnetic field in two adjacent minima. In particular, for
graphite again using for an estimate $\mu = 200 \mbox{K}$ and
$\Delta =0$ we obtain $|1/B_{i} - 1/B_{i+1}| = 8
\mbox{Tesla}^{-1}$. Writing the last equality in
Eq.~(\ref{period.graphite}), we introduced the area of an extremal
cross section of the Fermi surface, $A = \pi (\mu^2 - \Delta^2)$
as it is usually done. Thus, contrary to conventional systems with
quadratic dispersion law and the period of MO $\sim 1/\mu$, the
period of MO in the systems with linear dispersion (for $\Delta
=0$) is $\sim 1/\mu^2$ as seen in Fig.~\ref{fig:1}.

Making above estimates we have assumed everywhere that $\Delta =
0$. However, as it is clear from Eqs.~(\ref{period}) and
(\ref{period.graphite}), the opening of the gap $\Delta$ increases
the period of MO. In particular, for graphite even a rather small
value of $\Delta$, e.g. $\Delta \leq 10^{-2} \mu$, would produce a
sizeable $\sim 0.04 \mbox{Tesla}$ change in the period of MO. The
effect of the gap opening on the oscillations of the magnetization
is shown in Fig.~\ref{fig:2}. This figure was obtained using the
values of the parameters typical for graphite (see the caption).
As one can see, the gap opening produces an observable change in
the period of MO. Thus this method can be useful to test the
realization of magnetic catalysis phenomenon
\cite{Gusynin:1995:PRD} in graphite (see
Ref.~\cite{Gorbar:2002:PRB}), when the external magnetic field can
induce the opening of an insulating gap $\Delta(B)$ in the
relativistic spectrum of quasiparticle excitations. The crucial
condition for the observation of the $\delta B$ increase caused by
the gap opening is that the chemical potential $\mu$ itself {\it
does not change \/} \cite{Grigoriev:2001:JETP} as the gap $\Delta$
opens. Formally this condition corresponds to considering fixed
$\mu$ in the grand canonical ensemble. However, if the
experimental setup corresponds to the fixed carrier imbalance
$\rho$, one can see from Eqs.~(\ref{imbalance.T=0}) and
(\ref{mu-solution}) below that the entire difference $\mu^2 -
\Delta^2 = 2 \pi |\rho|/N$ adjusts to the strength of the applied
field, so that the period of MO remains unchanged, making the gap
detection impossible. Nevertheless, as we discuss at the end of
Sec.~\ref{sec:Omega.T}, there is still a possibility of the gap
detection if simultaneously with a frequency the amplitude of dHvA
oscillations is measured.

As pointed out in Ref.~\cite{Nguyen:2002:PRB}, the estimate of
$\delta B$ is not sufficient for the detectability of dHvA effect. In
addition we should calculate the magnetization and estimate the
magnitude of its oscillations.

\section{Magnetization}
\label{sec:magnetization}

The magnetization $M$ in the direction perpendicular to the plane
should be calculated in the canonical ensemble, thus we need the
Helmholtz free energy which is related to the thermodynamic
potential as
\be F(\rho,B)=\Omega(\mu(\rho,B),B)+\mu(\rho,B)\rho.
\label{free-energy}
\ee
Then \be \label{magnetization-def}
M_\rho(\rho,B)=-\frac{\partial F(\rho,B)}{\partial
B}=-\frac{\partial\Omega (\mu,B)}{\partial
B}\bigr|_{\mu=\mu(\rho,B)}=M_\mu(\mu(\rho,B),B), \ee where $M_\mu$
is the magnetization obtained in the grand canonical ensemble at
constant $\mu$:
\be
\label{magnetization}
M_\mu(\mu,B)=-\frac{\partial\Omega(\mu,B)}{\partial B}\bigr|_{\mu=
\mbox{const}},
\ee and we used that
$\partial\Omega(\mu,B)/{\partial\mu}=-\rho$. Formally, the
magnetization calculated in the canonical ensemble has the same
form as the magnetization calculated in the grand canonical
ensemble, but we have to take into account oscillations of $\mu$
in the system with fixed $\rho$. In the present paper we restrict
ourselves in what follows by studying MO in the limit $\omega_{c}
\ll \epsilon_F$, so that for fixed $\rho$, as shown in
Sec.~\ref{sec:number.Gamma}, one can neglect the oscillations of
$\mu(B)$.

The thermodynamic potential consists of the sum of three terms,
the vacuum energy (\ref{vacuum.energy}), regular
(\ref{Omega.T=0.reg}) and oscillating (\ref{Omega.T=0.osc}) parts.
We begin with the calculation of the part of the magnetization
which is due to the vacuum energy Eq.(\ref{vacuum.energy}).

As one could see from Eq.~(\ref{Omega.prime}), this energy
corresponds to the energy of the half-filled zone. Thus this case
is also interesting from physical point of view
\cite{Nersesyan:1989:JLTP}. Considering firstly the limit $\Delta
= 0$, we obtain from Eq.~(\ref{vacuum.energy.Delta=0}) that the
magnetization \be \label{magnetization.mu=Delta=0} M(\Delta
=0,\mu=0) = - \frac{3N\zeta(3/2)}{4 \sqrt{2} \pi^2} e^{3/2}
\sqrt{B} \ee (note the diamagnetic character of the vacuum
contribution in magnetization). The $\sim \sqrt{B}$ dependence of
the magnetization at $\mu =0$ implies that the susceptibility
$\chi =
\partial M/\partial B \varpropto B^{-1/2}$ diverges at zero field
raising a question about the stability of a homogeneous state
described by effective Lagrangian (\ref{Dirac.Lagrangian}). As
discussed in Refs.~\cite{Nersesyan:1989:JLTP,Nguyen:2002:PRB} in
the case of the OAF (DDW) this instability disappears if a
coupling between layers is included, or if finite
temperature and/or chemical potential is included. The instability
is also removed in the presence of the nonzero gap $\Delta$.
Indeed, using the asymptotics for zeta function \be \zeta(s,1+a)
\approx \frac{a^{1-s}}{s-1} -\frac{1}{2a^{s}}+\frac{s}{12a^{1+s}}
\ee for $a \gg 1$, we obtain from Eq.~(\ref{vacuum.energy}) that
\be M(\Delta, \mu =0) = -\frac{N e }{6\pi} \frac{e B}{\Delta},
\qquad eB \ll \Delta^2. \ee Thus the opening of the gap removes
the singularity at $B=0$ in the susceptibility.

Now we turn to the case $\mu > \Delta$ when in addition to the
vacuum contribution two more terms appear in the thermodynamic
potential, the regular $\Omega_{reg}(\mu)$ and oscillating
$\Omega_{osc}(\mu)$ parts, that at $T=0$ are given by
Eq.~(\ref{Omega.T=0.reg}) and Eq.~(\ref{Omega.T=0.osc}). As one
can easily see, for $\mu > \Delta$ field dependent terms in
Eq.~(\ref{Omega.T=0.reg}) coincide with the corresponding terms in
the vacuum energy (\ref{vacuum.energy}) up to a sign, so that
their total contribution to the magnetization is zero. Therefore,
only the  oscillating term of the thermodynamic potential given by
Eq.~(\ref{Omega.T=0.osc}) contributes to the magnetization.

In the small field limit, $e B \ll \mu^2$ differentiating the
potential (\ref{Omega.T=0.osc.Bernoulli}) we obtain
\be
\label{M.T=0.osc.Bernoulli}
M (\Delta,\mu) = - \frac{N e \mu
\theta(\mu - \Delta)}{2 \pi^{3/2}} \sum
\limits_{n=0}^\infty\frac{\Gamma(n+1/2)}{(n+1)!} \left[B_{n+2}
\left(\frac{w}{2} \right) - B_{n+1}\left(\frac{w}{2} \right)
\frac{w}{2} \right] \left( \frac{2e B}{\mu^2} \right)^{n+1}
\ee
where we used the relation $\partial B_{n}(z)/\partial z = n
B_{n-1}(z)$. Eq.(\ref{M.T=0.osc.Bernoulli}) is the generalization
to the relativistic spectrum of the formula for the magnetization
obtained by Shoenberg \cite{Shoenberg.book}.
For example, keeping first few terms in the expansion
(\ref{M.T=0.osc.Bernoulli}) we have
\be
M(\Delta =0, \mu) =
\frac{N e \mu}{2 \pi} \left[  B_1\left( \frac{v}{2} \right) -
\frac{3}{2} \frac{e B}{\mu^2 } B_2\left( \frac{v}{2} \right) -
\frac{1}{2} \frac{e^2 B^2}{\mu^4} B_3\left( \frac{v}{2} \right) +
O\left(\frac{e^3 B^3}{\mu^6} \right) \right],
\ee
where the
explicit expressions for the Bernoulli polynomials are given in
Eq.~(\ref{lowest.Bernoulli}). We recall that the Bernoulli
polynomials used here depend on the fractional part of their
argument so that magnetization is oscillating function with
the period $\delta B$ given by Eq.~(\ref{period}) and weakly
varying amplitude.
The dependence of the magnetization on the inverse field $B^{-1}$
is presented in Figs.~\ref{fig:1} and \ref{fig:2} computed on the
basis of Eq.~(\ref{M.T=0.osc.Bernoulli}).
The dependence $M(B^{-1})$ has a clear saw-tooth like shape showing that for
the considered range of the parameters, it is indeed convenient to
represent the oscillations in terms of the Bernoulli polynomials.

\section{Influence of impurities and temperature on magnetic oscillations}
\label{sec:Omega.Gamma-T}

\subsection{Thermodynamic potential $\tilde{\Omega}(\mu)$ and
equation for $\mu$ in the presence of scattering}
\label{sec:number.Gamma}

To obtain the equation for the chemical potential it is convenient
to use the DOS expressed as in Eq.~(\ref{DOS-through-logGamma}),
which allows one to write down the thermodynamic potential
(\ref{omega-subtracted.Gamma=0}), including the effect of impurity
scattering, as \be
\tilde{\Omega}_0(\mu)=-\frac{N\mu^2}{2\pi^2}\Gamma\ln\frac{\Lambda^2}
{2eB} -\frac{NeB}{\pi^2}\int\limits_0^\mu d\epsilon\,{\rm
Im}\left[\ln\Gamma
\left(\frac{\Delta^2-(\epsilon+i\Gamma)^2}{2eB}\right)+
\frac{1}{2}\ln\left(\frac{\Delta^2-(\epsilon+i\Gamma)^2}{2eB}\right)\right].
\ee Correspondingly, Eq.~(\ref{imbalance}) for the carrier
imbalance at $T=0$ and finite $\Gamma$ takes the form \be
\label{imbalance.T=0.Gamma}
\rho=-\frac{\partial\tilde{\Omega}_0(\mu)}{\partial\mu}
=\frac{N\mu}{\pi^2}\Gamma\ln\frac{\Lambda^2}{2eB}+\frac{NeB}{\pi^2}
{\rm Im}\left[\ln\Gamma\left(
\frac{\Delta^2-(\mu+i\Gamma)^2}{2eB}\right)+
\frac{1}{2}\ln\left(\frac{\Delta^2-(\mu+i\Gamma)^2}{2eB}\right)\right].
\ee For small $\Gamma \ll\mu, \Delta$, one can derive from
Eq.~(\ref{imbalance.T=0.Gamma}) a more simple expression \be
\label{imbalance.T=0} \epsilon_{F}^2= \mu^2+ \frac{2eB}{\pi}
\tan^{-1}\frac{ \sin [2\pi(\mu^2- \Delta^2)/(2eB)]}{\exp
[2\pi\mu\Gamma/(eB)]-\cos[2\pi(\mu^2- \Delta^2)/(2eB)]},
\quad\mu^2>\Delta^2, \ee where we have introduced the Fermi energy
$\epsilon_F$, counted from the edge of the gap $\Delta$, by means
of the relation $(N/2\pi)(\epsilon_F^2-\Delta^2) = \rho$. Note
that Eq.~(\ref{imbalance.T=0}) can also be derived directly from
the relationship $\rho= \int_0^\mu d\epsilon D_0(\epsilon)$ and
Eq.~(\ref{dos-oscil-gamma}) using the sum (\ref{sumformula1}).

In general, as  mentioned in the Introduction, both fixed $\mu$
and fixed $\rho$ cases are possible. For example, fixing $\mu$ in
Eq.~(\ref{imbalance.T=0.Gamma}) one can study the oscillations of
the carrier imbalance $\rho$ as a function of the field $B$.
Eq.~(\ref{imbalance.T=0}) can be considered as the relativistic
analog of the corresponding relation between $\mu$ and
$\epsilon_F$ in nonrelativistic case \cite{Champel:2001:PM}. One
can deduce formal "nonrelativistic" limit from
Eq.~(\ref{imbalance.T=0}) if one defines the nonrelativistic Fermi
energy $\epsilon_{F(nr)}= \pi\rho/N\Delta$ and corresponding
"nonrelativistic" chemical potential $\mu=\mu_{nr}+\Delta$ and
consider $\mu_{nr}<<\Delta$. This nonrelativistic limit is, in
fact, irrelevant for MO considered in the present paper, because
``relativistic'' theory studied here emerges as an effective
theory for nonrelativistic models described in
Sec.~\ref{sec:model}, so that it is meaningless to consider its
nonrelativistic limit.

\begin{figure}[h]
\centering{
\includegraphics[width=8cm]{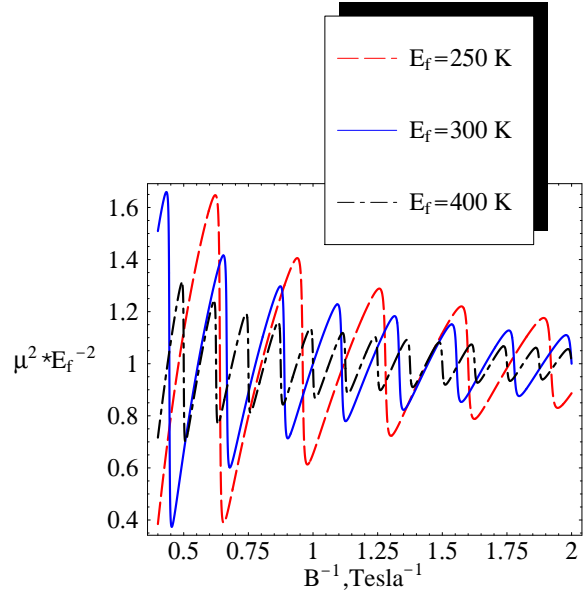}}
\caption{The chemical potential $(\mu/\epsilon_F)^2$ as a function
of inverse field, $B^{-1}$, for three different values of the
Fermi energy $\epsilon_F$, $\Gamma = 0.5 \mbox{K}$ and $T  =
\Delta = 0$. We use $eB \to 10^4 \mbox{K}^2 \cdot
B[\mbox{Tesla}]$.} \label{fig:3}
\end{figure}

On the other hand, for the magnetization  the case of a fixed
carrier imbalance $\rho$ is more natural, and therefore
oscillations of the chemical potential  have to be taken into
account by solving Eq.~(\ref{imbalance.T=0}) for $\mu(B)$.
Generally, this nonlinear equation cannot be solved analytically
without further approximations, but for $\epsilon_{F}^2 \gg e B$
the second term in Eq.(\ref{imbalance.T=0}) represents a small
correction to the constant value of the chemical potential so that
we obtain by iterations an approximate solution \be
\label{mu-solution} \mu^2 = \epsilon_F^2 - \frac{2eB}{\pi}
\tan^{-1}\frac{ \sin [2\pi (\epsilon_F^2 - \Delta^2)/(2eB)]} {\exp
[2\pi\epsilon_F\Gamma/(eB)]-\cos[2\pi(\epsilon_F^2 -
\Delta^2)/(2eB)]}. \ee It is plotted in Fig.~\ref{fig:3} for the
model parameters typical for graphite. As expected, these
oscillations have saw-tooth like shape and the sharpness of the
teeth depends on the value of $\Gamma$. One can also anticipate
that the relative amplitude of the oscillations decreases as the
Fermi energy $\epsilon_F$ increases and the field $B$ decreases.
This is indeed seen in Fig.~\ref{fig:3}. Yet in 2D the chemical
potential oscillations are rather strong even for large
$\epsilon_F$, due to the particular form of
Eq.~(\ref{imbalance.T=0}) that turns out to be different from its
3D counterpart.

In spite of the smallness of the second term in
Eq.~(\ref{mu-solution}), it is important to take into account the
difference of $\mu$ and $\epsilon_F$ in the arguments of
oscillating functions (in the magnetization, for example) in 2D,
because it alters the phase of oscillations. The difference can be
neglected if the additional condition $2 \pi \mu \Gamma \gtrsim
eB$ is valid (see discussion about the importance of the chemical
potential oscillations in Ref.~\cite{Champel:2001:PRB}). This
problem, however, is beyond the scope of the present work.

\subsection{Thermodynamic potential $\Omega_{osc}(\mu)$ and
magnetization $M_{osc}$ in the presence of scattering}
\label{sec:Omega.Gamma}

We return now to the oscillating part of the thermodynamic
potential (\ref{omega-subtracted.Gamma=0}) including the effect of
impurity scattering. The oscillating part can be written (see
Appendix~\ref{sec:C} for details) as \be \label{Omega.osc.Gamma}
\Omega_{osc}(\mu,\Gamma)=\frac{N(eB)^{3/2}}{2 \pi}\theta(\mu^2-
\Delta^2-\Gamma^2)\sum\limits_{k=1}^\infty \frac{1}{(\pi
k)^{3/2}}\left[ J_1\left(\pi kv, \pi k\gamma \right)\cos(\pi k w)+
J_2\left(\pi kv,\pi k\gamma\right)\sin(\pi kw)\right], \ee where
the functions of two arguments $J_1$(p,r) and $J_2(p,r)$ that
generalize the functions $J_1$ and $J_2$ defined in
Eq.~(\ref{J.def}),  are defined in Eq.~(\ref{J.new}) and the
variables $w$, $u$ and $\gamma$ are \be \label{w-u.Gamma}
w=\frac{\mu^2-\Delta^2-\Gamma^2}{eB},\quad
u=\frac{\Delta^2+\Gamma^2}{eB}, \quad \gamma =
\frac{\Gamma^2}{eB}. \ee Note that the variable $w$ (and also $u$)
that was defined in Sec.~\ref{sec:Omega0.Gamma=0} by
Eq.~(\ref{w-v}) is redefined in the presence of scattering in
accordance with Eq.~(\ref{w-u.Gamma}). As before the frequency of
MO is given by $w/2$, so that it would diminish in the presence of
impurities if we considered fixed $\mu$ case. This decrease of the
MO frequency for fixed $\mu$ is specific for the present
relativistic model, while in the nonrelativistic case the MO
frequency remains intact even in the presence of impurities. The
origin of the difference between relativistic and nonrelativistic
cases can be traced back to the difference in zero field DOS for
these cases.

For  $\Gamma,\sqrt{eB}<<\mu$ keeping the leading term in
asymptotic expansions (\ref{J1}), (\ref{J2}) (this is the first
term in Eq.~(\ref{J1}) only), we obtain from
Eq.~(\ref{Omega.osc.Gamma}) (omitting theta function) \be
\label{omega-oscil-gamma}
\Omega_{osc}(\mu,\Gamma)=\frac{N(eB)^2}{2\pi\mu}\sum\limits_{k=1}^\infty
\frac{\cos(\pi kw)}{(\pi k)^2}\exp\left({-2\pi
k\frac{\mu\Gamma}{eB}}\right), \quad \mu >0. \ee It is easy to
check that for $\Gamma=0$ Eq.~(\ref{omega-oscil-gamma}) reduces to
the expression that follows directly from
Eq.~(\ref{Omega.T=0.osc}) if one substitutes  the first term of
the expansion (\ref{J_int_asym}), $J_{1}(p) \approx 1/\sqrt{p}$.
This comparison with Eq.~(\ref{Omega.T=0.osc}) allows to identify
in Eq.~(\ref{omega-oscil-gamma}) the {\em Dingle factor} \be
\label{Dingle.factor} R_D(k, \mu) = \exp \left(-2\pi k
\frac{\mu\Gamma}{eB}\right), \ee which describes reduction of the
amplitude of $k$-th harmonic due to the electron scattering. As is
seen, the finite amount of impurities leads to smoothing of the
oscillations. The factor (\ref{Dingle.factor}) is a generalization
of the Dingle harmonic damping \be \label{Dingle.factor.nonrel}
R_D^{\rm NR}(k) \varpropto \exp\left(-2\pi
k\frac{\Gamma}{\omega_c}\right), \ee for the nonrelativistic case,
usually written in terms of the nonrelativistic cyclotron
frequency $\omega_c$. Comparing Eqs.~(\ref{Dingle.factor}) and
(\ref{Dingle.factor.nonrel}), one can notice two facts. First one
was already discussed in Sec.~\ref{sec:GF}, viz. in the
relativistic case $R_D$ can be larger than it would be for the
nonrelativistic case for the same field for $\mu \lesssim  m v_F
v_2 = \omega_L^2/(2 \omega_c)$, so that the condition for the
observation of MO in the relativistic case is less strict. In
particular, for graphite this condition remains valid for $\mu
\lesssim m_e v_F^2 \approx 6 \times 10^4 \mbox{K}$. Secondly,
further increase of $\mu$ makes MO unobservable, contrary to the
nonrelativistic case, when $R_D$ does not depends on the chemical
potential $\mu$.

Calculating the magnetization (\ref{magnetization}) and keeping
only dominant terms, we obtain from Eq.~(\ref{omega-oscil-gamma})
\be
\begin{split}
\label{magnetization.Gamma}
M_{osc}& =-\frac{N(\mu^2-\Delta^2-\Gamma^2)}{2\pi\mu}
\sum\limits_{k=1}^\infty\frac{\sin(\pi kw)}{\pi k}
e^{-2\pi k\frac{\mu\Gamma}{eB}}\\
& =- \frac{N(\mu^2-\Delta^2-\Gamma^2)}{2\pi\mu}
\tan^{-1}\frac{ \sin [2\pi(\mu^2-\Delta^2-\Gamma^2)/(2eB)]}
{\exp [2\pi \mu\Gamma/(eB)]-\cos[2\pi(\mu^2-\Delta^2-\Gamma^2)/(2eB)]},
\end{split}
\ee
where in the second equality we used Eq.~(\ref{sumformula1}). We will show now
that the oscillating part of the magnetization $M_{osc}$ is directly
related to the oscillating part of the chemical potential.

Comparing the result Eq.(\ref{magnetization.Gamma}) for $M_{osc}$
with the solution for chemical potential (\ref{mu-solution}), one
can notice that the oscillating part of the magnetization is
directly proportional to the oscillating part of $\mu^2$: \be
M_{osc}=\frac{N(\epsilon_F^2-\Delta^2)}{4
eB\epsilon_F}(\mu^2)_{osc} =\frac{\pi
\rho}{2eB}\frac{(\mu^2)_{osc}}{\epsilon_F} \ee (this is obvious if
one neglects small terms $\Gamma^2$ in the trigonometric functions
in Eq.~(\ref{magnetization.Gamma})). This proportionality can be
in fact seen when Fig.~\ref{fig:3} (plotted for finite $\Gamma$)
is compared with Figs.~\ref{fig:1} and \ref{fig:2} (plotted for
$\Gamma =0$). Note that unlike nonrelativistic case
\cite{Champel:2001:PM} the magnetization is proportional to
$\mu^2_{osc}$, not  $\mu_{osc}$. If the ratios $\epsilon^2_F/eB$
and $2 \pi \mu \Gamma/eB$ are small, the oscillating part of the
chemical potential under trigonometric functions in
Eq.(\ref{magnetization.Gamma}) leads to additional frequencies in
the Fourier spectrum of the magnetization. In its turn this
changes the harmonics amplitudes compared with the LK like formula
Eq.(\ref{magnetization.Gamma}). In the non-relativistic case this
was demonstrated in
Ref.~\cite{Champel:2001:PRB,Grigoriev:2001:JETP,Alexandrov:1996:PRL}.

\subsection{Calculation of $\Omega_{osc} (\mu)$ for $T \neq 0$}
\label{sec:Omega.T}

Now we consider the effect of finite temperature
on the oscillations of the thermodynamic potential that
can be calculated from the zero temperature
potential (\ref{Omega.osc.Gamma}) using the convolution
property Eq.~(\ref{thermal_convolution}).
For that it is convenient to rewrite the expression (\ref{Omega.osc.Gamma})
in the form
\begin{equation}
\Omega_{osc}(\mu,\Gamma)= - \frac{N(eB)^{3/2}}{2\pi^{3/2}}
\theta(\mu^2-\Delta^2-\Gamma^2)\sum\limits_{k=1}^\infty
\frac{1}{(\pi k)^{3/2}}{\rm Im}\left[e^{-\frac{i\pi}{4}}\int\limits_0^\infty
\frac{dt\,e^{-i(\pi kv)t+\frac{i\pi k\gamma}{t}}}{\sqrt{t}(t+1)}e^{-i\pi k w}
\right],
\end{equation}
where we used that the functions $J_1(p,r)$ and $J_2(p,r)$ can be represented
as $-\mbox{Im}$ and $\mbox{Re}$ parts of the same function
\be
\sqrt{\pi}J_1(p,r)=- {\rm Im}\int\limits_0^\infty\frac{dt\,e^{-pt-r/t}}
{\sqrt{t}(t+i)},\quad \sqrt{\pi}J_2(p,r)= {\rm Re}\int\limits_0^\infty
\frac{dt\,e^{-pt-r/t}}{\sqrt{t}(t+i)},
\ee
and rotated the integration contour to the imaginary axis. Hence
\be
\begin{split}
\Omega_{osc}(T,\mu,\Gamma)&= -\frac{N(eB)^{3/2}}{2\pi^{3/2}}
\sum\limits_{k=1}^\infty
\frac{1}{(\pi k)^{3/2}}{\rm Im}\left[e^{-\frac{i\pi}{4}}\int\limits_0^\infty
\frac{dt\,e^{i\pi k\gamma/t}}{\sqrt{t}(t+1)}\int\limits_{-\infty}^\infty
\frac{d\epsilon}{4T\cosh^2\frac{\epsilon-\mu}{2T}}\right. \\
&\times \left.\theta(\epsilon^2-\Delta^2
-\Gamma^2)\exp\left(-\frac{it\pi k\epsilon^2}{eB}-i\pi k\frac
{\epsilon^2-\Delta^2-\Gamma^2}{eB}\right)\right].
\end{split}
\ee At low temperatures, $T\to0$, after making a shift
$\epsilon\to\epsilon+\mu$ and changing the variable of integration
$\epsilon\to 2T\epsilon$, the integral over $\epsilon$ gives
(neglecting $T^2$ term in the exponent) \be
\int\limits_{-\infty}^\infty\frac{d\epsilon}{\cosh^2\epsilon}e^{-\frac{i\pi
k} {eB}(t+1)4T\mu\epsilon}=
2\int\limits_{0}^\infty\frac{d\epsilon}{\cosh^2\epsilon}\cos\left(
\frac{4T\mu(t+1)\pi
k}{eB}\epsilon\right)=\frac{4\pi^2kT\mu(t+1)}{eB\sinh
\frac{2\pi^2kT\mu(t+1)}{eB}}, \ee where we used the formula
(3.982.1) from Ref.~\cite{Gradshtein.book}. We obtain \be
\label{Omega.T-Gamma}
\begin{split}
\Omega_{osc}(T,\mu,\Gamma)&= -\frac{N(eB)^{1/2}T\mu}{\sqrt{\pi}}
\theta(\mu^2-\Delta^2-\Gamma^2)\sum\limits_{k=1}^\infty
\frac{1}{(\pi k)^{1/2}}\\
&\times{\rm Im}\left[e^{-\frac{i\pi}{4}}e^{-i\pi kw}
\int\limits_{0}^\infty\frac{dt}{\sqrt{t}}e^{-\frac{i\pi
kt\mu^2}{eB}+ \frac{i\pi
k\gamma}{t}}\frac{1}{\sinh\frac{2\pi^2kT\mu(t+1)}{eB}}\right].
\end{split}
\ee Nonoscillating factor in the integrand has a maximum at $t=0$,
thus we can take $\sinh$ evaluated at $t=0$ while the remaining
integral over $t$ is evaluated by means of
Eqs.~(\ref{mcdonalds-representation}),
(\ref{mcdonalds-halfinteger}). We finally get (omitting again
theta function) \be \label{Omega.osc.T}
\begin{split}
\Omega_{osc}(T,\mu,\Gamma)& =NeBT\sum\limits_{k=1}^\infty
\frac{\cos(\pi kw)}{\pi k}\frac{1}{\sinh\frac{2\pi^2kT\mu}{eB}}
\exp \left[-2\pi k\frac{\mu\Gamma}{eB}\right] \\
& = \frac{N(eB)^2}{2\pi\mu}\sum\limits_{k=1}^\infty
\frac{\cos(\pi kw)}{(\pi k)^2} R_T(k,\mu) R_D(k,\mu),
\end{split}
\ee where we introduced the {\it temperature amplitude factor} \be
\label{temperature.factor} R_T(k,\mu) = \frac{2\pi^2 k
T\mu/(eB)}{\sinh\frac{2\pi^2kT\mu}{eB}}, \ee which is a
``relativistic'' equivalent of the temperature reduction factor of
the famous LK formula. Clearly, since $R_T (k,\mu) \to 1$ for $T
\to 0$, Eq.~(\ref{Omega.osc.T}) reduces to
Eq.~(\ref{omega-oscil-gamma}) and for finite $T$ the thermal
broadening causes a reduction of the MO amplitude with respect to
$T =0$ case. This becomes more evident if we consider the limit $B
\to 0$, so that the function $1/\sinh [2 \pi^2 k T \mu/(eB)] \sim
\exp[- 2 \pi^2 k T \mu/(eB)]$. Comparison of this exponent with
Eq.~(\ref{Dingle.factor}) allows one to introduce the {\em Dingle
temperature \/} \be \label{Dingle.temperature} T_D =
\frac{\Gamma}{\pi}, \ee so that a reduction of amplitude due to
quasiparticles scattering from impurities can be interpreted as
leading to an effective rise of  a temperature from true
temperature $T$ to $T+T_D$, i.e., as if the system could not be
cooled below a minimal temperature $T_D$. Note that while the
Dingle factors (\ref{Dingle.factor}) and
(\ref{Dingle.factor.nonrel}) are different, the value of $T_D$
itself is the same for relativistic and nonrelativistic cases.
Comparing Eq.~(\ref{Omega.osc.T}) with the corresponding
expression (15) for $\Omega_{osc}$ in 2D from
Ref.~\cite{Champel:2001:PM} (this expression contains also a spin
factor $R_s$ which is not considered in our work), one can notice
that relativistic and nonrelativistic cases differ only by the
factor $(-1)^k$ that appears due to the presence of the zero
energy, $E_0 = \omega_c/2$ for the nonrelativistic Landau levels.

It is seen from Eq.~(\ref{Omega.osc.T}) that the conditions for
observation of dHvA oscillations can be formulated as an
inequality for the strength of magnetic field: \be
\label{condition} \frac{\mu^2-\Delta^2}{2} \gtrsim eB \gtrsim {\rm
max}\{2\pi^2T\mu,2\pi^2T_D\mu \}. \ee The first inequality is
necessary in order to have at least one oscillation. It implies
physically that the field must not be so high that all fermions
are in the lowest Landau level, i.e. the filling factor has to be
bigger than one. The second inequality guarantees that the
amplitude of oscillations is not exponentially suppressed, and for
that the magnetic field must be strong enough to make spacing
between Landau levels bigger or of the order of the thickness of
the thermal layer times the Fermi energy.

\begin{figure}[h]
\centering{
\includegraphics[width=8cm]{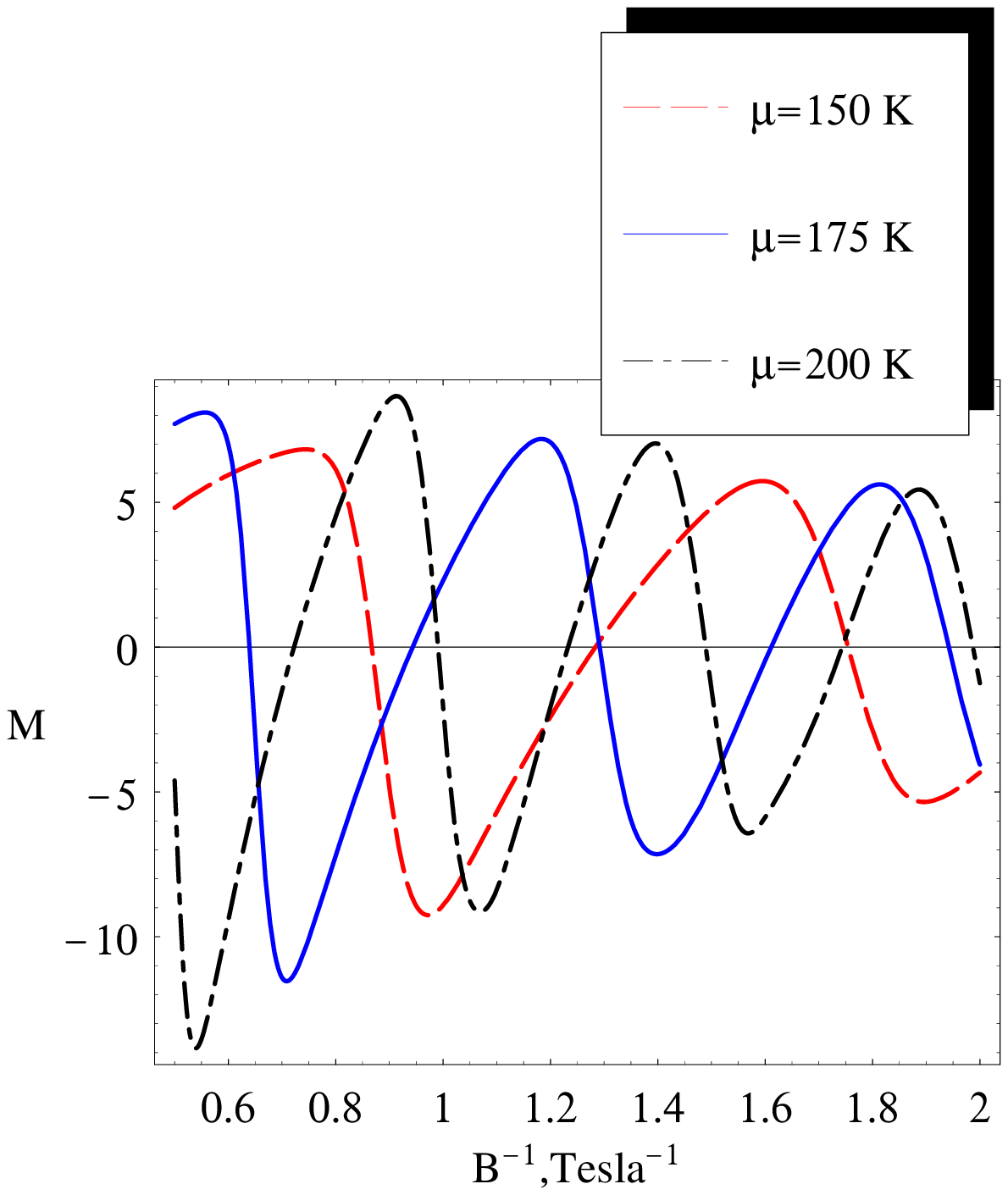}}
\caption{The magnetization $M$ (in $\mbox{K}/(\mbox{Tesla cm}^2)$
calculated for $N=1$) as a function of inverse field, $B^{-1}$,
for three different values of the chemical potential. We use $eB
\to 10^4 \mbox{K}^2 \cdot B[\mbox{Tesla}]$, $\Gamma = 0.5
\mbox{K}$, $T = 5 \mbox{K}$ and $\Delta =0$.} \label{fig:4}
\end{figure}

\begin{figure}[h]
\centering{
\includegraphics[width=8cm]{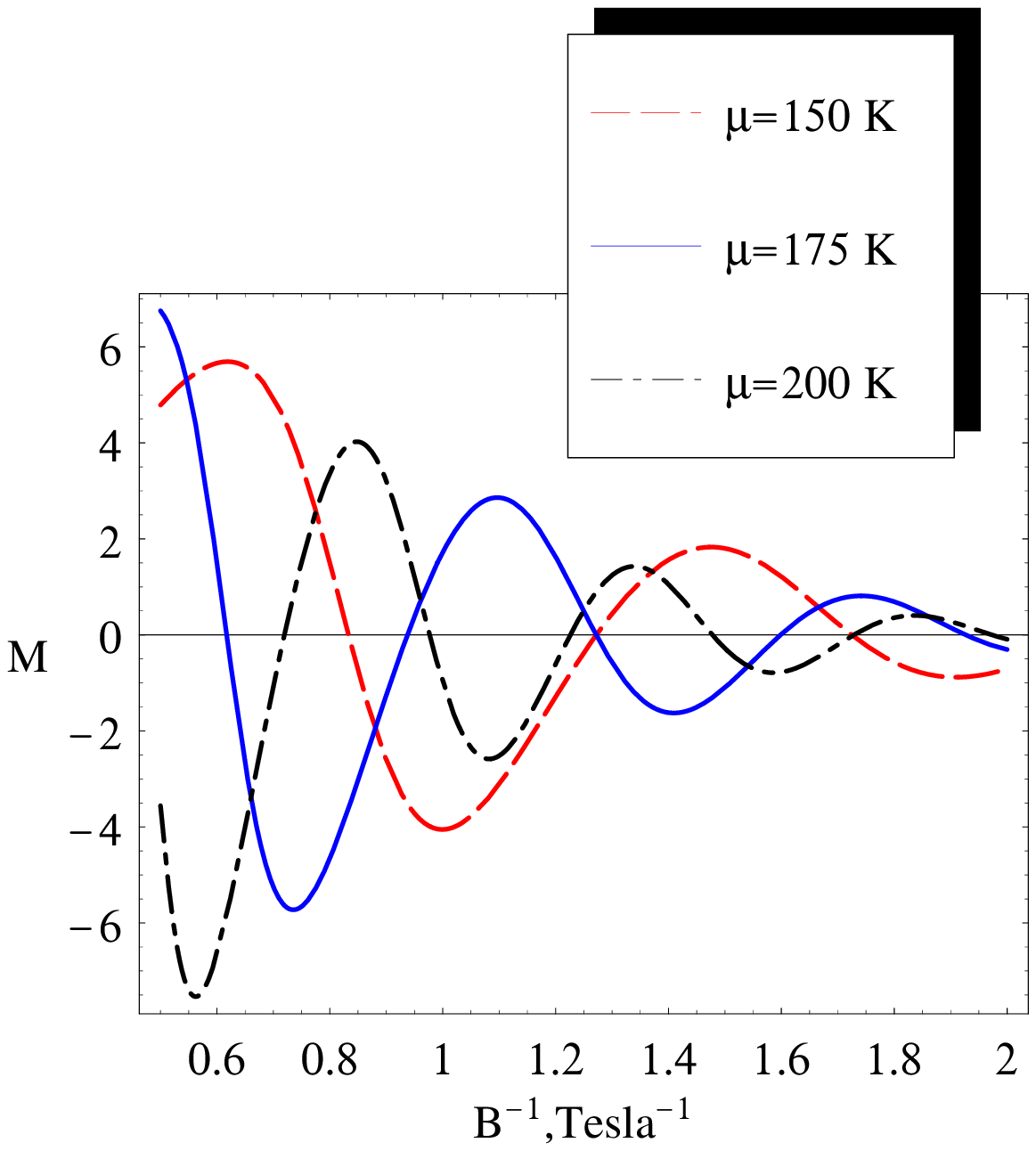}}
\caption{The magnetization $M$ (in $\mbox{K}/(\mbox{Tesla cm}^2)$
calculated for $N=1$) as a function of inverse field, $B^{-1}$,
for three different values of the chemical potential. We use $eB
\to 10^4 \mbox{K}^2 \cdot B[\mbox{Tesla}]$, $\Gamma = 0.5
\mbox{K}$, $T = 15 \mbox{K}$ and $\Delta=0$.} \label{fig:5}
\end{figure}

The dependence of the magnetization (\ref{magnetization}) on
$B^{-1}$ is obtained by differentiation of Eq.~(\ref{Omega.osc.T})
and the results of numerical computations are shown in
Figs.~\ref{fig:4} and \ref{fig:5}. They illustrate the above
mentioned facts that the MO are smoothed as the temperature and/or
the chemical potential rise. Obviously the increase of $\Gamma$
gives the same result. Moreover, one can clearly see the damping
of dHvA amplitude as the field $B$ decreases. It is known that the
damping of dHvA oscillation amplitude as function of several
parameters is commonly used to extract system parameters such as
Dingle temperature and the effective electron mass. However, these
methods of analysis depend on the applicability of the LK theory.
Assuming that the relativistic generalization of the LK theory
considered here is valid at least for $\omega_L \ll \epsilon_F$,
we suggest that, using Eqs.~(\ref{magnetization.Gamma}) and
(\ref{Omega.osc.T}), one can extract both the value of $\mu$ from
the damping and the difference, $\mu^2 - \Delta^2$ from the
frequency of dHvA oscillations. The knowledge of these two
parameters allows to obtain the value of the gap $\Delta$ even for
the fixed carrier imbalance $\rho$.

\section{Conclusions}
\label{sec:conclusions}

In this paper we have studied magnetic oscillations in
thermodynamic quantities such as DOS, thermodynamic potential and
magnetization in planar systems with the relativistic Dirac-like
spectra for quasiparticle excitations. The attention was mainly
paid to the regime, where the calculations in canonical (fixed
$\rho$) and grand canonical (fixed $\mu$) ensembles give
equivalent results. Our main results can be summarized as follows.

\noindent (1) We have obtained analytical expressions that
describe MO in the DOS given by Eqs.~(\ref{DOS.oscillations}),
(\ref{DOS.Re-Im})  for zero width $\Gamma =0$ (no impurities) and
Eq.~(\ref{dos-oscil-gamma}) for finite width $\Gamma$ for Landau's
levels.

\noindent (2) For $\Gamma =0$, we have expressed the MO in
thermodynamic potential, Eq.~(\ref{Omega.T=0.osc.Bernoulli}) and
magnetization, Eq.~(\ref{M.T=0.osc.Bernoulli}) as a series in the
periodically continued Bernoulli polynomials. This representation
turns out to be useful for 2D case when the oscillations have
saw-tooth like shape.  In the limit $\omega_L \ll \mu$ it is
sufficient to consider only first few terms of the series.

\noindent (3) For finite impurity scattering rate $\Gamma$
and temperature  we have obtained the thermodynamic potential
(\ref{Omega.osc.T}) and represented it in terms of the Dingle factor
$R_D$, Eq.~(\ref{Dingle.factor}), and the temperature amplitude factor
$R_T$, Eq.~(\ref{temperature.factor}),  as it is
usually done in  Lifshits-Kosevich theory. The spin factor, $R_s$, can
be calculated similarly if the Zeeman term is included in the model.

\noindent (4) For finite $\Gamma$, we have derived also the equation
(\ref{imbalance.T=0.Gamma}) for the chemical potential. Its
solution for fixed carrier imbalance exhibits oscillations of
$\mu$ as a function of the magnetic field. It is shown that the
oscillating part of the magnetization is in fact directly
proportional to the oscillating part of the chemical potential
(more precisely, to $(\mu^2)_{osc}$).

\noindent (5) On the basis of obtained formulas, we have discussed
the possibility to detect a gap that may open in the spectrum of
the Dirac-like quasiparticle excitations due to a nontrivial
interaction between them.

One of the possible and necessary extensions of our results would
be to study magnetic oscillations for the case when the
consideration within the canonical ensemble is crucial, i.e. in
the regime where the oscillations of the chemical potential
described by Eq.~(\ref{imbalance}) at fixed $\rho$ ($\epsilon_F$)
cannot be neglected. The necessity of such an extension emerges
from the fact that the oscillations of $\mu$ become important for
$2 \pi \mu \Gamma \lesssim e B$ which is exactly the condition
(\ref{condition}) for having large amplitude of the MO.

Another important extension would be to consider MO analytically
in the transport quantities such as electrical and thermal
conductivities. These oscillations were in fact seen in the
numerical results presented in Ref.~\cite{Sharapov:2003:PRB}, but
analytical results can be useful for comparison with the
experiments \cite{Uji:1998:PB,Wang:2003:PLA}. While for dHvA
effect the condition $\rho = \mbox{const}$ is more natural, it is
plausible that SdH effect can be measured under condition $\mu =
\mbox{const}$. This, as we have discussed in the paper, can be
used for an experimental observation of the gap $\Delta$,
especially if the gap depends on the applied field, $\Delta =
\Delta(B)$.

All above-mentioned problems can be treated analytically due to
the fact that the broadening of Landau levels has a Lorentzian
shape and the impurity scattering rate $\Gamma$ (the width of this
distribution) is assumed to be field and temperature independent.
In fact, both these assumptions can be questioned, in particular
for 2D systems with a linear dispersion. For example, the validity
of the first assumption for 2D systems is now discussed in the
literature (see Refs. in \cite{Champel:2001:PRB}). While the
second assumption may well be valid in the low field regime, it is
necessary to investigate its domain of validity and probably to
consider also the dependence $\Gamma(B)$ \cite{Champel:2002:PRB}
to access the high-field regime.

\section{Acknowledgments}
We gratefully acknowledge N.~Andrenacci and V.M.~Loktev for
stimulating discussions. One of us, S.G.Sh., thanks P.~Esquinazi
for a correspondence. This work was supported by the research
project 2000-067853.02/1 of the Swiss NSF. V.P.G. acknowledges
support from the Natural Sciences and Engineering Research Council
of Canada. The work of V.P.G. was supported also by the
SCOPES-projects 7UKPJ062150.00/1 and 7 IP 062607 of the Swiss NSF.

\appendix
\section{Calculation of vacuum energy $\Omega_0(\mu=0,B)$}
\label{sec:A}

As mentioned above, the vacuum contribution does not change under
averaging over thermal and $\Gamma$ distributions, so it is
sufficient to calculate it at $T =0$ and in the absence of
impurities.  We calculate the vacuum term using the density of
states in the form of Eq.~(\ref{DOS.delta})
\begin{equation}
\Omega_0(\mu=0,B)= -\int\limits_{0}^{\infty} d\epsilon\epsilon
D_0(\epsilon)=-\frac{NeB}{2\pi}\left[\Delta+2\sum\limits_{n=1}^\infty
M_n\right].
\end{equation}
 The sum over the Landau levels is divergent and to calculate it
 we use the representation
\begin{equation}
\frac{1}{a^s}=\frac{1}{\Gamma(s)}\int\limits_0^\infty
dtt^{s-1}e^{-at}
\end{equation} to write
\begin{equation}
\Omega_0(\mu=0,B)=-\frac{N
eB}{2\pi\Gamma(-1/2)}\int\limits_0^\infty\frac{dt}{t^{3/2}}\,
e^{-t\Delta^2}\left[1+2\sum\limits_{n=1}^\infty e^{-2eBt}\right]=
\frac{N}{4\pi^{3/2}}\int\limits_{1/\Lambda^2}^\infty\frac{dt}{t^{5/2}}\,
e^{-t\Delta^2}eBt\coth(eB t),
\end{equation}
 where we introduced the ultraviolet cutoff $\Lambda$ as the bandwidth.
The integral can be evaluated through the generalized zeta function
\be
\label{vacuum.energy}
\Omega_0(\mu=0,B)=-\frac{N}{2\pi}\left[\frac{\Lambda\Delta^2}{\sqrt{\pi}}
+\Delta eB + (2eB)^{3/2}\zeta\left(-\frac{1}{2},\frac{\Delta^2}{2eB}+1\right)
\right]+O\left(\frac{1}{\Lambda}\right),
\ee
where the term $\sim\Lambda^3$ which does not depend on
$\Delta$ and $B$ is omitted. The last
expression coincides with the vacuum energy computed in the second
paper in Ref.~\cite{Gusynin:1995:PRD}. Setting in Eq.~(\ref{vacuum.energy})
$\Delta=0$ and using the identity
$\zeta(-1/2,1)=\zeta(-1/2)= -(1/4\pi)\zeta(3/2)$ we obtain
\be
\label{vacuum.energy.Delta=0}
\Omega_0(\mu=0,B, \Delta =0)=
\frac{N \zeta(3/2)}{2 \sqrt{2}\pi^{2}} (eB)^{3/2}.
\ee
To compare Eq.~(\ref{vacuum.energy.Delta=0})
with Eq.~(15) of Ref.~\cite{Nersesyan:1989:JLTP} and
Eq.~(10) of Ref.~\cite{Nguyen:2002:PRB}, one should
restore the prefactor $1/(v_F v_D)$ with
the velocities $v_F$, $v_{D}$ and express them via the parameters
$t$ and $D_0$ as done below Eq.~(\ref{Hamiltonian.DDW.matrix}).
Note that the ground state energy increases with applied
magnetic field, indicating diamagnetism.

\section{Calculation of $\tilde{\Omega}_0(\mu)$}
\label{sec:B}

Substituting Eq.~(\ref{DOS.osc-first}) in Eq.~(\ref{omega-subtracted.Gamma=0})
and integrating by parts we obtain
\be
\tilde{\Omega}_0(\mu)=-\frac{N}{2\pi}\int\limits_0^\mu d\epsilon
\theta(\epsilon^2-\Delta^2)\left[\epsilon^2-\Delta^2
+2eB\sum\limits_{k=1}^\infty\frac{1}{\pi k}\sin\frac{\pi k(\epsilon^2
-\Delta^2)}{eB}\right]\equiv \Omega_0^{(1)}(\mu)+\Omega_0^{(2)}(\mu),
\ee
where
\be
\Omega_0^{(1)}(\mu)=-\frac{N}{6\pi}\theta(\mu-\Delta)(\mu-\Delta)^2
(\mu+2\Delta),
\ee
and
\be
\label{Omega.2}
\begin{split}
\Omega_0^{(2)}(\mu)&=-\frac{NeB}{\pi} \theta (\mu - \Delta)
\sum\limits_{k=1}^\infty\frac{1}
{\pi k}\int\limits_\Delta^\mu d\epsilon\sin\frac{\pi k(\epsilon^2
-\Delta^2)}{eB}\\
&=-\frac{N(eB)^{3/2}}{2\pi} \theta (\mu - \Delta)
\sum\limits_{k=1}^\infty\frac{1}
{\pi k}\int\limits_0^w \frac{dx\sin(\pi kx)}{\sqrt{x+u}}.
\end{split}
\ee
Writing the last expression for
$\Omega_0^{(2)}(\mu)$ we introduced new variables
$w$ defined in Eq.~(\ref{w-v}) and  $u=\Delta^2/(eB)$.

To extract explicitly the oscillations in the variable $w$ we rewrite the
integral in Eq.~(\ref{Omega.2}) as
\be
\label{I-integral}
\begin{split}
&\int\limits_0^w \frac{dx\sin(\pi kx)}{\sqrt{x+u}}=\int\limits_0^w
d x \sin(\pi kx)\frac{1}{\sqrt{\pi}}\int\limits_0^\infty\frac{dt}{\sqrt{t}}
e^{-t(x+u)}\\
&=\frac{1}{\pi\sqrt{k}}\left[\int\limits_0^\infty\frac{dt\,e^{-t(\pi ku)}}
{\sqrt{t}(t^2+1)}-\int\limits_0^\infty\frac{dt\,e^{-t(\pi kv)}}
{\sqrt{t}(t^2+1)}\cos(\pi k w)-\int\limits_0^\infty\frac{dt\,\sqrt{t}
e^{-t(\pi kv)}}{t^2+1}\sin(\pi k w)\right],
\end{split}
\ee
where $v$ is defined in Eq.~(\ref{w-v}).
The integrals in the last equation can be expressed in terms of the degenerate
hypergeometric function $\Psi$:
\be
\label{J.def}
\begin{split}
\sqrt{\pi}J_1(p)&= \int\limits_0^\infty\frac{dt\,e^{-tp}}{\sqrt{t}(t^2+1)}=-
{\rm Im}\int\limits_0^\infty\frac{dt\,e^{-tp}}
{\sqrt{t}(t+i)}=-\sqrt{\pi}\,{\rm Im}\left[i^{-1/2}\Psi
\left(\frac{1}{2},\frac{1}{2};ip\right)\right]=-\sqrt{\pi}\,{\rm Im}
\left[\sqrt{p}\,\Psi\left(1,\frac{3}{2};ip\right)\right],\\
\sqrt{\pi}J_2(p)& = - \frac{\partial}{\partial p}
[\sqrt{\pi}J_1(p)]=
 \int\limits_0^\infty\frac{dt\,\sqrt{t}e^{-tp}}
{t^2+1}={\rm Re}\int\limits_0^\infty\frac{dt\,e^{-tp}}
{\sqrt{t}(t+i)}=\sqrt{\pi}\,{\rm Re}\left[\sqrt{p}\,\Psi\left(1,
\frac{3}{2};ip\right)\right],
\end{split}
\ee
where we used also the relationship \cite{Gradshtein.book}
\be
\Psi(a,b;z)=z^{1-b}\Psi(a-b+1,2-b;z).
\ee
One can readily see that the integrals $J_1,J_2$ are monotonic functions of their
arguments and have the following asymptotic expansions
\be
\label{J_int_asym}
J_1(p)=\frac{1}{\sqrt{\pi}}\sum\limits_{n=0}^\infty\frac{(-1)^n
\Gamma(2n+1/2)}{p^{2n+1/2}}, \quad
J_2(p)=\frac{1}{\sqrt{\pi}}\sum\limits_{n=0}^\infty\frac{(-1)^n
\Gamma(2n+3/2)}{p^{2n+3/2}}.
\ee Thus for regular part of the potential we obtain \be
\label{Omega.T=0.mon}
\Omega_{reg}(\mu)=-\frac{N}{6\pi}\theta(\mu-\Delta)(\mu-\Delta)^2
(\mu+2\Delta)-\frac{N(eB)^{3/2}}{2 \pi} \theta(\mu - \Delta)
\sum\limits_{k=1}^\infty\frac{J_1(\pi ku)} {(\pi k)^{3/2}}, \ee
while the oscillating part takes the form (\ref{Omega.T=0.osc})
written in Sec.~\ref{sec:Omega0.Gamma=0}.
Eqs.~(\ref{Omega.T=0.mon}) and (\ref{Omega.T=0.osc}) were also
derived  in Ref.~\cite{Vshivtsev:1996:JETP}  using a different
approach \cite{foot1}. The sum in Eq.~(\ref{Omega.T=0.mon}) can be
evaluated through the generalized zeta function in the following
way \be
\begin{split}
&\sum\limits_{k=1}^\infty\frac{J_1(\pi ku)}{(\pi k)^{3/2}}=
\frac{1}{\sqrt{\pi}}\sum\limits_{k=1}^\infty
\frac{1}{(\pi k)^{3/2}}\int\limits_0^\infty
\frac{dt\,e^{-\pi kut}}{\sqrt{t}(t^2+1)}=\frac{1}{\sqrt{\pi}}
\int\limits_0^\infty\frac{dt\,e^{-ut}}{\sqrt{t}}\sum\limits_{k=1}^\infty
\frac{1}{t^2+(\pi k)^2}\\
&=\frac{1}{2\sqrt{\pi}}\int\limits_0^\infty dt\,t^{-3/2}e^{-ut}
\left(\coth t-\frac{1}{t}\right)=-2^{3/2}\zeta\left(-\frac{1}{2},
1+\frac{u}{2}\right)-u^{1/2}-\frac{2}{3}u^{3/2}
\end{split}
\ee which finally results in Eq.~(\ref{Omega.T=0.reg}).

The oscillating part $\Omega_{osc}(\mu)$ of the thermodynamic potential
can be also represented in terms of the generalized zeta function. For
that one should use $\Gamma(n+\alpha)=\int_0^\infty ds s^{n+\alpha-1}e^{-s}$
and perform summation over $n$ by means of the formula
\be
\sum\limits_{n=0}^\infty\frac{x^n}{n!}B_n(y)=\frac{xe^{xy}}{e^x-1},
\quad |x|<2\pi,
\ee
we obtain
\be
\sum\limits_{n=2}^\infty\frac{\Gamma(n+\alpha)B_n(y)}{n!}x^n=\int
\limits_0^\infty ds\,s^{\alpha-1}e^{-s}\left[\frac{sx\,e^{sxy}}
{e^{sx}-1}-1-sxB_1(y)\right].
\ee
Using now the formula \cite{Gradshtein.book}
\be
\int\limits_0^\infty\frac{x^{\nu-1}e^{-\mu x}dx}{1-e^{-\beta x}}=\frac{1}
{\beta^\nu}\Gamma(\nu)\zeta\left(\nu, \frac{\mu}{\beta}\right),\quad
{\rm Re}\mu>0, {\rm Re}\nu>0,
\ee
we find
\be
\begin{split}
\Omega_{osc}(\mu)=&-\frac{N\mu^3}{2\pi}\theta(\mu-\Delta)
\left\{\left(\frac{2eB}{\mu^2}\right)^{3/2}
\zeta\left(-\frac{1}{2},1+\frac{\mu^2}{2eB}-{\rm mod}\left[\frac{\mu^2
-\Delta^2}{2eB}\right]\right)\right.\\
&+\left.\frac{2}{3}-\frac{2eB}{\mu^2}\left( {\rm
mod}\left[\frac{\mu^2-\Delta^2}{2eB}\right]-\frac{1}{2}\right)\right\}.
\end{split}
\ee
This expression involves a dependence on the nonanalytical
fractional part ${\rm
mod}\left[\frac{\mu^2-\Delta^2}{2eB}\right]$. In terms of the
integer part $\left[\frac{\mu^2-\Delta^2}{2eB}\right]$ the total
thermodynamic potential (a regular plus oscillating parts) can be
written as \cite{Cangemi:1996:AP}
\be \label{poten-through-zeta}
\begin{split}
\tilde{\Omega}_0(\mu)=
-\frac{NeB}{\pi}\theta(\mu-\Delta)&\left\{\mu\left[\frac{\mu^2
-\Delta^2}{2eB}\right]+\sqrt{2eB}\zeta\left(-\frac{1}{2},1+\frac{\Delta^2}
{2eB}  + \left[\frac{\mu^2-\Delta^2}{2eB}\right]\right) \right.\\
& + \left. \frac{\mu-\Delta}{2}
-\sqrt{2eB}\zeta\left(-\frac{1}{2},1+\frac{\Delta^2}{2eB}\right)\right\}.
\end{split}
\ee
The last two terms are in fact cancelled by the zero chemical potential
(vacuum) part of
the thermodynamic potential Eq.~(\ref{vacuum.energy}) (for
$\mu>\Delta$).

In fact, the expression Eq.~(\ref{poten-through-zeta}) can be
obtained much easier if one uses Eq.~(\ref{DOS.delta}) and writes
Eq.~(\ref{Omega.T=0.osc}) as \be \label{Omega.sum}
\tilde{\Omega}_0(\mu)=-\frac{NeB}{2\pi}\left(\mu-\Delta+\sum\limits_{n=0}^{\left
[\frac{\mu^2-\Delta^2}{2eB}\right]}(\mu-M_n)\right). \ee Using now
the formula for the generalized zeta function \be
\zeta(z,v+k)=\zeta(z,v)-\sum\limits_{n=0}^{k-1}(n+v)^{-z} \ee one
immediately arrives at  Eq.~(\ref{poten-through-zeta}). However,
the oscillating property of $\tilde{\Omega}_0(\mu)$ is not so
transparent in Eq.~(\ref{poten-through-zeta}) as it is in
Eqs.~(\ref{Omega.T=0.osc}) and (\ref{Omega.T=0.osc.Bernoulli}). On
the other hand, the expression for $\tilde{\Omega}_0(\mu)$ through
the generalized zeta-function is convenient for studying large
field  behavior. Indeed, for $eB>>\Delta^2, \mu^2-\Delta^2$ we
find from Eq.(\ref{poten-through-zeta}) that in this limit \be
\tilde{\Omega}_0(\mu)\simeq -\frac{NeB}{2\pi}\theta(\mu-\Delta)
(\mu-\Delta), \ee which is precisely the contribution due to the
lowest Landau level which is the only one to survive in the high
field limit. Also, as is seen from Eq.(\ref{poten-through-zeta}),
the oscillating term (first zeta-function in curve brackets) is
necessary to take into account even in the high field regime in
order to cancel $(eB)^{3/2}$ term and obtain the correct linear
term for $\tilde{\Omega}_0(\mu)$.

\section{Calculation of $\Omega_{osc}(\mu)$ in the presence of scattering}
\label{sec:C}

It turns out that due to the bad convergence of the integrals like
(\ref{Omega.PG}) and (\ref{Omega.PG-PT}) with the Lorentzian
distribution $P_\Gamma(\omega)$ this calculation can be done more
easily if we directly substitute the  DOS (\ref{dos-oscil-gamma})
in  Eq.~(\ref{omega-subtracted.Gamma=0}) and generalize the
calculation made in Sec.~\ref{sec:Omega0.Gamma=0}. Choosing again
for definiteness $\mu >0$, for the  oscillating part of the
thermodynamic potential we have
\begin{equation}
\begin{split}
\Omega_{osc}(\mu,\Gamma)&= -\frac{NeB}{\pi}\theta(\mu^2-\Delta^2-\Gamma^2)
\sum\limits_{k=1}^\infty\frac{1}
{\pi k}\int\limits_{\sqrt{\Delta^2+\Gamma^2}}^\mu d\epsilon
\sin\frac{\pi k(\epsilon^2-\Delta^2-\Gamma^2)}{eB}\exp\left(
-\frac{2\pi k|\epsilon|\Gamma}{eB}\right) \\
&=-\frac{N(eB)^{3/2}}{2\pi}\theta(\mu^2-\Delta^2-\Gamma^2)
\sum\limits_{k=1}^\infty\frac{1} {\pi k}\int\limits_0^w
\frac{dx\sin(\pi kx)}{\sqrt{x+u}} \exp\left(- 2\pi k
\sqrt{\gamma(x+u)}\right),
\end{split}
\end{equation}
where the variables $w$, $u$ and $\gamma$ are
defined in Eq.~(\ref{w-u.Gamma}).
Using the representation
\begin{equation}
\frac{e^{-z}}{z}=\frac{1}{2\sqrt{\pi}}\int\limits_0^\infty
\frac{dt\, e^{-t-\frac{z^2}{4t}}}{t^{3/2}},
\end{equation}
we can perform the integration over $x$ to get
\begin{equation}
\begin{split}
& \int\limits_0^w\frac{dx\sin(\pi kx)}{\sqrt{x+u}}
\exp\left(-2\pi k\sqrt{\gamma(x+u)}\right) \\
&=\frac{1}{\pi\sqrt{k}}\int\limits_0^\infty\frac{dt}{\sqrt{t}(t^2+1)}
e^{-\frac{\pi k\gamma}{t}} \left[e^{-t(\pi ku)}-\cos(\pi k w)
e^{-t(\pi kv)}-t\sin(\pi k w)e^{-t(\pi kv)}\right],
\end{split}
\end{equation}
where  the variable $v= \mu^2/eB$ is the same  as
in Eq.~(\ref{w-v}).
Hence the oscillating part of the thermodynamic potential can be
written (omitting the monotonic term related to the first term in
square brackets of the last expression) in
the final form (\ref{Omega.osc.Gamma})
using the functions of two arguments $J_1$(p,r) and $J_2(p,r)$
\be
\label{J.new}
\sqrt{\pi}J_1(p,r)=\int\limits_0^\infty\frac{dt\,e^{-tp-\frac{r}{t}}}
{\sqrt{t}(t^2+1)},\quad
\sqrt{\pi}J_2(p,r) = -\frac{\partial}{\partial p}[ \sqrt{\pi}J_1(p,r)] =
\int\limits_0^\infty\frac{dt\,\sqrt{t}\,
e^{-tp-\frac{r}{t}}}{t^2+1}.
\ee
As is seen, the functions
$J_1(p,r),J_2(p,r)$ are monotonic functions of their variables.

For large values of the parameter $p$ we can obtain asymptotic expansion
of the functions $J_1(p,r),J_2(p,r)$ changing the variable $t\to t/p$
and expanding then the denominator in powers of $p$, we get
\be
\begin{split}
\sqrt{\pi}J_1(p,r)&=\sum\limits_{n=0}^\infty\frac{(-1)^n}{p^{2n+1/2}}
\int\limits_0^\infty\frac{dt\,
e^{-t-\frac{rp}{t}}}{t^{-2n+1/2}}=2\sum\limits_{n=0}^\infty(-1)^n
\left(\frac{r}{p}\right)^{n+1/4}K_{2n+1/2}(2\sqrt{rp}),
\end{split}
\ee where we used the formula for the representation of the
Macdonald function \be
K_\nu(2z)=\frac{z^\nu}{2}\int\limits_0^\infty\frac{dt\,e^{-t-\frac{z^2}{t}}}
{t^{\nu+1}}, \qquad {\rm Re}z^2>0, \quad |{\rm
arg}z|<\frac{\pi}{2}. \label{mcdonalds-representation}
\ee
and the
relation $K_{-\nu}(z)=K_{\nu}(z)$. Since for half integer values
of the index $\nu$
\be
K_{n+\frac{1}{2}}(z)=\sqrt{\frac{\pi}{2z}}e^{-z}\sum\limits_{k=0}^{n}
\frac{(n+k)!}{k!(n-k)!(2z)^k} \label{mcdonalds-halfinteger}
\ee
(see Eq.(8.468) in \cite{Gradshtein.book}), we obtain for the first
few terms in the expansion
\be \label{J1}
J_1(p,r)\simeq\frac{1}{\sqrt{p}}\left[1-\frac{3}{4p^2}\left(1 + 2
\sqrt{rp}+\frac{4}{3}rp\right)\right]e^{-2\sqrt{rp}},\quad
p\to\infty.
\ee
Similarly, for $J_2$ we get \be \label{J2}
\begin{split}
\sqrt{\pi}J_2(p,r)&= 2\sum\limits_{n=0}^\infty(-1)^n
\left(\frac{r}{p}\right)^{n+3/4}K_{2n+3/2}(2\sqrt{rp})\\
&\simeq\frac{1}{p^{3/2}}\left[\frac{1}{2}+\sqrt{rp}-
\frac{1}{p^2}\left(\frac{15}{8}+\frac{15}{4}\sqrt{rp}+3rp+(rp)^{3/2}
\right)\right]e^{-2\sqrt{rp}},\quad p\to\infty.
\end{split}
\ee

\newpage

\end{document}